\documentclass[journal,comsoc]{IEEEtran}
\pdfoutput=1

\usepackage[T1]{fontenc}% optional T1 font encoding
\ifCLASSINFOpdf
\else
\fi

\usepackage[cmintegrals]{newtxmath}
\usepackage{cite}
\usepackage{amsmath}
 
\usepackage{amssymb,amsfonts}
\usepackage{algorithmic}
\usepackage{graphicx}
\usepackage{textcomp}
\usepackage{xcolor}
\usepackage{caption}
\usepackage{multirow}
\definecolor{blue}{rgb}{0,0,1} 
\begin{document}

\title{Large-scale Machine Learning Cluster Scheduling via Multi-agent Graph Reinforcement Learning}
%
%
% author names and IEEE memberships
% note positions of commas and nonbreaking spaces ( ~ ) LaTeX will not break
% a structure at a ~ so this keeps an author's name from being broken across
% two lines.
% use \thanks{} to gain access to the first footnote area
% a separate \thanks must be used for each paragraph as LaTeX2e's \thanks
% was not built to handle multiple paragraphs
%

\author{Xiaoyang~Zhao,~\IEEEmembership{Student Member,~IEEE,}
        Chuan~Wu,~\IEEEmembership{Senior Member,~IEEE} \\
        Department of Computer Science, The University of Hong Kong, Email: \{xyzhao, cwu\}@cs.hku.hk
        \thanks{This work was supported in part by grants from Hong Kong RGC under the contracts HKU 17204619, 17208920, 17207621 and C5026-18G (CRF).}
}
        
% X. Zhao and C. Wu are with the Department of Computer Science, the University of Hong Kong, Hong Kong (e-mail: xyzhao@cs.hku.hk, cwu@cs.hku.hk).

% The paper headers
% \markboth{IEEE TRANSACTIONS ON NETWORK AND SERVICE MANAGEMENT}%
% {Shell \MakeLowercase{\textit{et al.}}: Bare Demo of IEEEtran.cls for IEEE Communications Society Journals}

% make the title area
\maketitle

% As a general rule, do not put math, special symbols or citations
% in the abstract or keywords.
\begin{abstract}
Efficient scheduling of distributed deep learning (DL) jobs in large GPU clusters is crucial for resource efficiency and job performance. While server sharing among jobs improves resource utilization, interference among co-located DL jobs occurs due to resource contention. Interference-aware job placement has been studied, with white-box approaches based on explicit interference modeling and black-box schedulers with reinforcement learning. 
In today's clusters containing thousands of GPU servers, running a single scheduler to manage all arrival jobs in a timely and effective manner is challenging, due to the large workload scale. We adopt multiple schedulers in a large-scale cluster/data center, and propose a multi-agent reinforcement learning (MARL) scheduling framework to cooperatively learn fine-grained job placement policies, towards the objective of minimizing job completion time (JCT). 
To achieve topology-aware placements, our proposed framework uses hierarchical graph neural networks to encode the data center topology and server architecture. In view of a common lack of precise reward samples corresponding to different placements, a job interference model is further devised to predict interference levels in face of various co-locations, for training of the MARL schedulers. Testbed and trace-driven evaluations show that our scheduler framework outperforms representative scheduling schemes by more than 20\% in terms of average JCT, and is adaptive to various machine learning cluster topologies.
\end{abstract}

% Note that keywords are not normally used for peerreview papers.
\begin{IEEEkeywords}
AI for Management, Cluster Scheduling, Distributed Machine Learning, Multi-agent Reinforcement Learning, Graph Neural Network.
\end{IEEEkeywords}

\IEEEpeerreviewmaketitle

\section{Introduction}
% 1.5 page
Deep Learning (DL) is powering various services today, spreading across fields including computer vision, language processing, recommendation systems, etc. Learning a modern DL model over large datasets (aka a DL job) is usually carried out using a distributed training framework such as MXNet~\cite{example} or Tensorflow~\cite{tensorflow}, and run using multiple GPUs.

Nowadays many leading IT companies operate large machine learning (ML) clusters, where a large number of DL jobs are run to learn various ML models. 
Cluster schedulers are responsible for producing scheduling policies for DL workloads in such a cluster, whose decisions are crucial for efficient utilization of the very expensive hardware resources and for model learning expedition. For example, workers and parameter servers (PS) of a distributed job (using the PS architecture) can be distributed onto different servers based on the placement decisions of schedulers, in the case that they cannot be entirely hosted on the same server due to resource fragmentation or the purpose of load balancing.
An efficient placement policy for these tasks could expedite training and increase cluster throughput, especially in a large-scale cluster with various DL workloads.

The main issue often neglected by many existing schedulers~\cite{borg}\cite{mesos} is that, while server sharing among co-located DL jobs improves resource utilization, inter-job interference occurs due to resource contention. Within the same server, jobs share resources such as PCIe bus (for frequent communication between CPU and GPU), QPI (for inter socket between CPUs), I/O and CPU cores; across the servers, jobs share network bandwidth and switch capacity. Jobs training different DL models have unique resource characterization. For example, training VGG16~\cite{vgg16} demands more of network I/O for communicate gradients; CTC~\cite{ctc} is CPU intensive due to word embedding generations. This implies that, different co-locations of jobs will lead to different levels of interference, which encourages cluster operators to co-locate jobs with low levels of interference to maximize training performance.

The potential of interference-aware scheduling has been discussed in the literature. Some white-box studies build explicit interference models to predict performance slowdown of co-locations, e.g., for MapReduce tasks~\cite{mapreduce}, VM tasks~\cite{tracon} with I/O contention, etc. Other DNN-based approaches use a large amount of historical trace to learn interference levels of co-located ML jobs~\cite{gexin}, or equip the scheduler with a reinforcement learning (RL) model to improve job placement policy through explorations and feedback~\cite{harmony}\cite{springer-rl}.

Purely white-box solutions heavily rely on the precision of the performance model, which often requires deep dive into each application's execution and careful optimizations of coefficients. Generality will be an issue, as such heuristics lack adaptability to various workload types, server configurations, and data center topologies. RL schedulers do not require detailed analytical interference modeling, but often face the scalability issue: in a large-scale DL cluster with thousands of GPU servers, learning to schedule heavy workloads with one RL scheduler is time-demanding and likely not converging to good policies.

% That is, in a large-scale DL cluster with thousands of GPU servers, applying a single scheduler to learn to schedule heavy workload brings up challenges: 

% (i) the huge dimension of input state prevent agent from learning valuable features; (ii) the growing action space requires a long-term exploration, which could make the model trap into local optimal or even non-converge; (iii) the scheduler needs to perform a huge number of inferences for all jobs submitted to the cluster, both in training and execution phase, thus leading to low training speed and extra scheduling delay. \xyzhao{simplify, move RL to motivation}

In addition, cross-machine communication is important for the performance of distributed training jobs. An effective scheduling policy should be aware of the link states and topology structures. For example, congested links should be avoided for gradient synchronization with strategical task placement in a DL job. Existing schedulers~\cite{eks}~\cite{tetris} used in production clusters are largely topology-oblivious.

%worker and PS are supposed to be placed as close as possible if available resource allows, and 

Aiming at mitigating inter-job interference and reducing communication overhead, we design a scalable job scheduling framework for large-scale DL clusters, adopting multiple RL schedulers to cooperatively decide the placement of newly submitted jobs. Each scheduler in our framework is a learning agent that encodes its local cluster topology and workload information, to learn placement policies that minimize average job completion time through exchanging observations with other schedulers. Specifically, we make the following contributions in developing this scheduling framework:

$\triangleright$ We verify the inefficiency of representative locality-preserving and congestion-unawareness job placements, and analyze the interference levels of different jobs in case of different co-locations. Instead of using heuristic interference models to achieve interference avoidance, we adopt a multi-agent reinforcement learning (MARL) model to schedule DL workloads, achieving better generality to unknown environments (with interference situation not experienced before), as well as ensure the scalability of the scheduler. 

$\triangleright$ We design a hierarchical Graph Neural Network (GNN) with edge information encoded to capture the cluster/data center topology, server architectures and link status. Following the GNN hierarchy, the schedulers aggregate local observations by exchanging with other schedulers, and make topology-aware job placement decisions.

$\triangleright$ We also quantify the performance slowdown of co-located DL jobs through an interference model. The interference model addresses a common lack of precise performance samples corresponding to different placements, needed during the offline training phase of RL schedulers. With the inference model, schedulers are able to learn a fine-grained placement policies, in terms of which GPU group (i.e., GPUs connected to the same CPU) to place each worker/PS of a job onto.

$\triangleright$ We conduct extensive trace-driven evaluation to compare our framework with representative scheduling schemes under different workloads and cluster settings. Results show that our solution achieves at least 20\% improvement in terms of average job completion time, at a scale of thousands of GPU servers with various configurations. As compared to using a single RL scheduler, MARL achieves better policy convergence and faster policy learning speed. We also implement a prototype on Kubernetes, which further verifies the effectiveness of proposed framework.

\section{Background and Motivation}
% 1.5 page
% challenges in MARL
\subsection{Graph Convolutional Network}
While traditional deep learning (e.g., using CNN, RNN) captures the hidden representation of Euclidean data, there is an increasing number of applications
where data are represented in the form of non-Euclidean graphs. For example, the data center network can be modeled as a graph, where servers and switches are linked to each other via transmission paths. Recently, deep learning approaches have been extended to such graph data. Similar to 2D convolution, one may perform graph convolutions by taking the weighted average of features in a node's neighborhood to achieve feature aggregation~\cite{gnn}.

\vspace{-3mm}
\subsection{Distributed Training}
ML model training usually minimizes a loss function over a large dataset (e.g., ImageNet~\cite{example}), which is time and resource consuming. Hence, many DL frameworks have been designed for distributed training, %Currently, most distributed training DL frameworks (
e.g., MXNet~\cite{example}, Tensorflow~\cite{tensorflow}.  %) employ the parameter server architecture.
% 

%In distributed training with PS architecture, the DNN model is partitioned among multiple parameter servers and the training data is split among workers.
In a data-parallel distributed training job, each worker computes the parameter gradients locally using one mini-batch from allocated training data, exchanges gradients with other workers and updates the global model accordingly. One epoch of training refers to processing the entire training dataset once. The training proceeds for multiple epochs until model convergence (i.e., when loss function converges or accuracy reaches a threshold) or the total number of epoches reaches a predefined maximum epoch number.

\vspace{-3mm}
\subsection{Inter-Job Interference}
%To mitigate cross-machine communication overhead, 
For resource efficiency, a common practice is to maximally pack jobs onto the servers. For example, Google Borg~\cite{borg} and Tetris~\cite{tetris} adopt multi-resource bin packing, under which workloads will be consolidated on the least number of machines to reduce resource fragments, and to utilize servers effectively.
However, performance interference exists among co-located jobs, as jobs with the same burst demand of resources may possibly be co-located together. 

\begin{figure}
\centering
\includegraphics[width=0.48\textwidth]{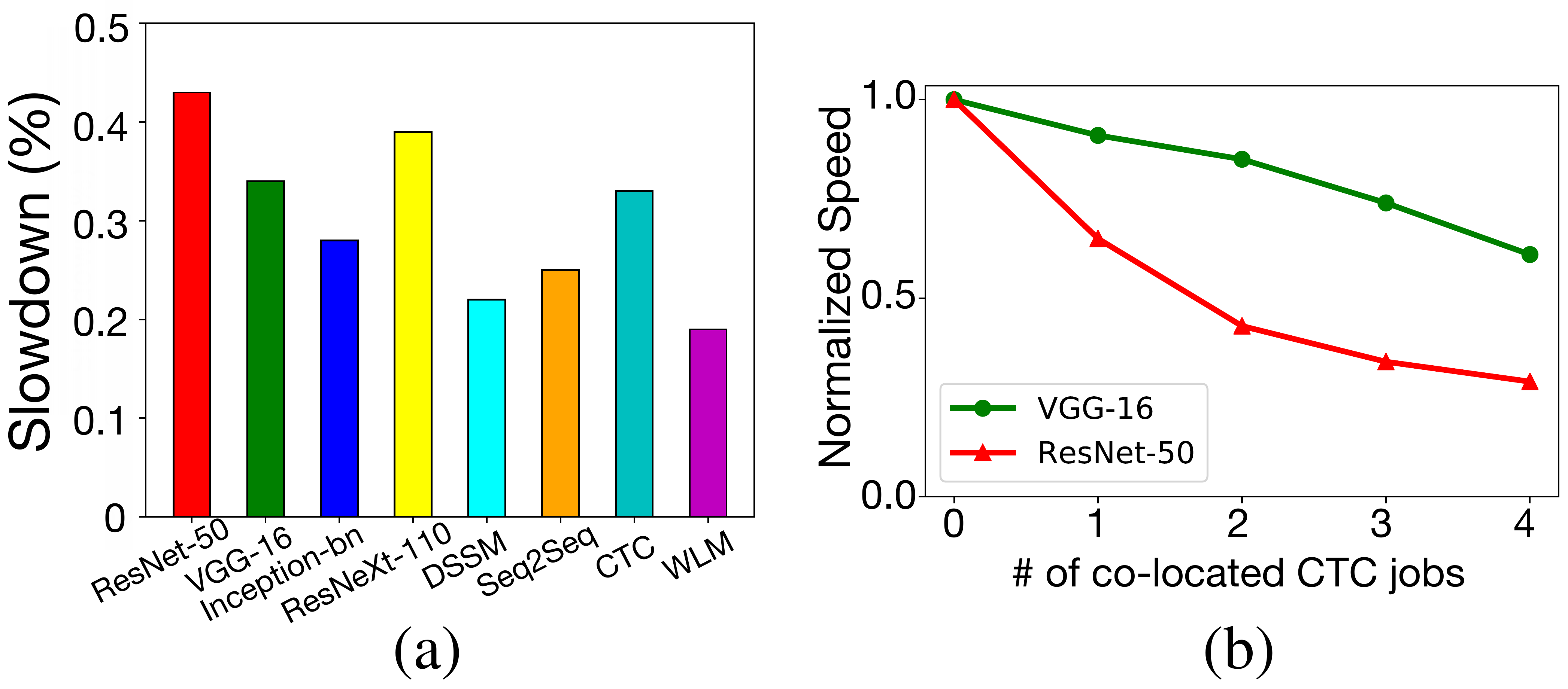}
\caption{Inter-job interference: (a) Performance slowdown using bin-packing scheme, as compared to standalone execution; (b) Normalized speed with increasing number of CTC jobs.}  % and a customer of no one.}
\label{fig:motivation-1}
\end{figure}

%The 8 ML models are designed for different application domain with unique NN architectures and resource utilization. 
We show interference among different ML model training jobs (Table~\ref{ml-table}), by comparing the training speed of these jobs when packed onto 4 servers using multi-resource bin packing, to standalone execution (i.e., each job is run on
a dedicated server).
Each job demands 6 CPU cores and 1 GPU; each server is equipped with one Intel E5-1660 CPU with 16 cores and two GTX 1080Ti GPUs. We observe a 30\% training slowdown on average in Fig.~\ref{fig:motivation-1}(a). The interference occurs due to contention of resources among co-located jobs. For example, although heavily relying on GPUs to accelerate computation, training a DL job still requires CPU processing and frequent communication between GPU and CPU through the shared PCIe bus.

%This performance degradation will be more serious in a large-scale cluster with heavy workload, since a bad placement of one PS/worker will slowdown a synchronous training job.

% ml table, add citation if space allows
\begin{table}[]
\small
\centering
\caption{DL jobs with different model types, implemented according to official MXNet tutorials~\cite{example}}\label{ml-table}
\begin{tabular}{|c|c|c|}
\hline
Model                   & Application domain      & Dataset  \\ \hline
ResNet-50           & image classification    & ImageNet \\
VGG-16              & image classification    & ImageNet \\
Inception-bn        & image classification    & Caltech  \\
ResNeXt-110         & image classification    & CIFAR10  \\
DSSM                & word representation     & text8    \\
Seq2Seq             & machine translation     & WMT17    \\
CTC                 & sentence classification & mr       \\
WLM                 & language modeling       & PTB      \\ \hline
\end{tabular}
\end{table}

% The interference occurs due to contention of resources among co-located jobs. For example, a DL job relies on GPU to accelerate computation, but it still requires frequent communication with CPU through the shared PCIe bus. 
We notice that different levels of interference occurs when jobs training different types of models are co-located. This is because each model architecture has its own resource sensitivity. Such job characterization leads to unique reaction of model training performance to the change of available resources (i.e., when a new job arrives or an existing job is completed on a server). To verify this, we increase the number of CTC training jobs on one GPU of a server, and record the training speed of co-located ResNet50 and VGG16 on the same server but different GPUs, respectively. We can observe from Fig.~\ref{fig:motivation-1}(b) that ResNet50 suffers from worse performance degradation due to greater contention of CPU resources (virtual cores and cache), which it relies on more.

\begin{figure}
\centering
\includegraphics[width=0.48\textwidth]{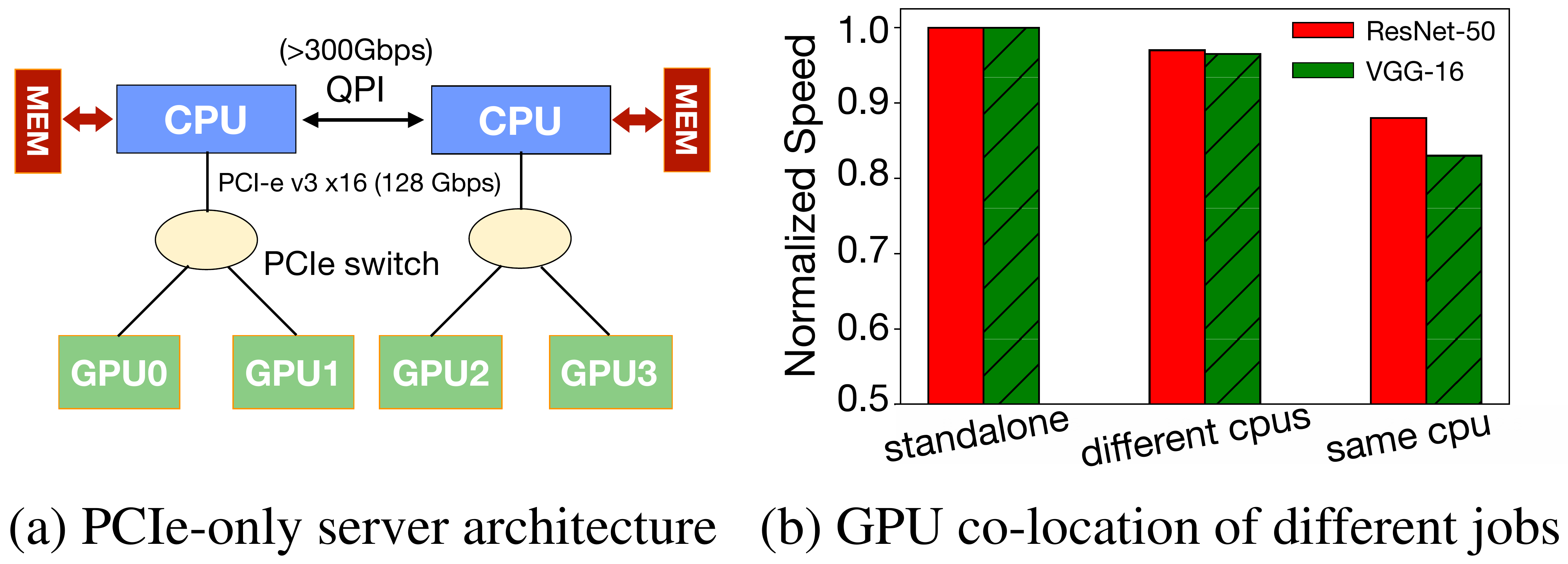}
\caption{Inter-job interference relates to placement within server.}% and a customer of no one.}
\label{fig:motivation-2}
\end{figure}

\vspace{-5mm}
We further identify that the inter-job interference is related to the specific server architecture (e.g., a PCIe-only system is shown in Fig.~\ref{fig:motivation-2}(a)). In Fig.~\ref{fig:motivation-2}(b), we compare two jobs (each using 1 GPU) running on the same server with different GPU locality settings (on different CPUs and on the same CPU), and also compare with the baseline of running each job solely on the server (the `standalone' case in Fig.~\ref{fig:motivation-2}(b)). We observe that the performance drops as the GPUs (where the two jobs are run) are using the same CPU, which leads to more competition of shared PCIe bus and CPU resources. This encourages us to design fine-grained placement policy, taking server architecture into consideration.

\subsection{Impact of Communication}
One natural idea to reduce resource competition is to apply load balancing %scheme 
(or similar variants) as adopted in Mesos~\cite{mesos} and Kubernetes~\cite{eks}, by assigning each worker/PS to the server with the least load. As compared to fully locality-based schemes, load balancing avoids high resource utilization and reduces interference level implicitly. However, this advantage is no longer obvious in a cluster under heavy workloads, due to the cross-machine parameter synchronization communication overhead and data center bandwidth contention. %that results. 

\begin{figure}
\centering
\includegraphics[width=0.48\textwidth]{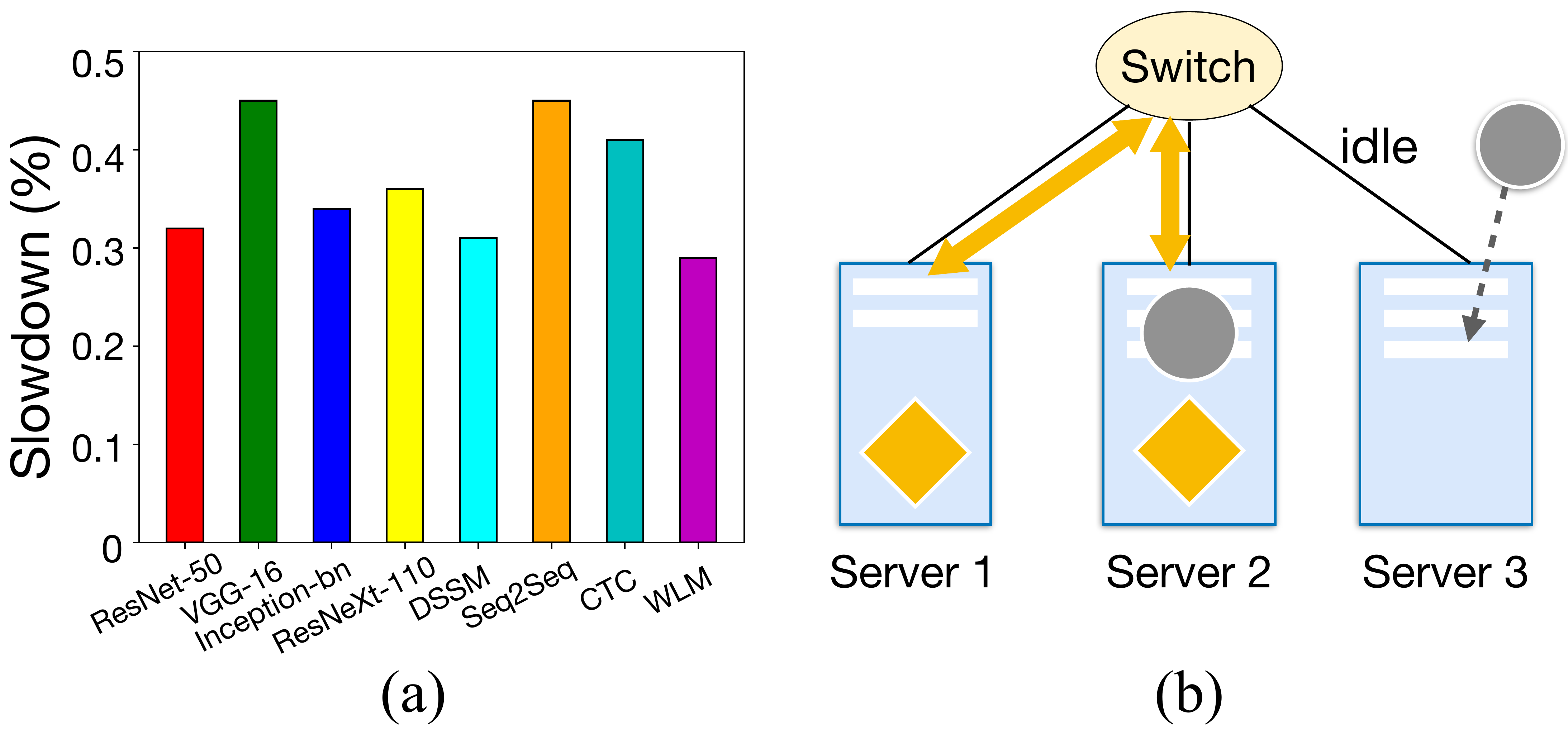}
\caption{Impact of communication: (a) Performance slowdown using load-balancing scheme, as compared to standalone execution; (b) Placement of 2 jobs %(denoted use different shapes)
onto 3 servers. Yellow arrows represent network traffic. % and suppose that ``diamond" job and ``triangle" job are training similar model.
}% and a customer of no one.}
\label{fig:motivation-3}
\end{figure}

To verify the inefficiency of load balancing, we compare the training performance when placing the 8 ML jobs from Table~\ref{ml-table} onto 4 servers using this scheme, to the jobs' respective standalone execution on a single server. We observe a 37\% training slowdown on average, as shown in Fig.~\ref{fig:motivation-3}(a). Even for jobs such as training ResNet50, whose computation phase accounts for a large portion of one mini-batch processing time, performance degradation is still serious due to the impact of extra communication when the workers and PSs are placed across different servers.
 
%In the case that workers/PSs of one distributed job are not able to be packed within the same server (due to potential interference or resource fragment), an optimal placement policy should pack them as close as possible, which suppose the scheduler to capture the cluster topology. 
Other than maximally colocating tasks of the same job within the same server(s), which brings about potential interference as we have mentioned before, communication overhead could be mitigated if the scheduler has a global view of both the cluster topology and link status. We use three servers and one switch as shown in Fig.~\ref{fig:motivation-3}(b) for an example. ``Circles'' denote tasks (PS/worker) of one job, and ``diamonds'' denote another job. Each server can maximally host two tasks. One task of the ``circle'' job has been placed on Server 2, which also holds one task of the other job. The other task of the ``circle" job can be placed on Server 1 or Server 3, and Server 3 is a better choice due to lower communication traffic along the routing path between Server 2 and Server 3. % which increases the burden on traversing links and introduces extra delay. 
In a production cluster shared by a large number of jobs, cross-machine traffic is common; hence the awareness of link status counts when generating scheduling policies. Unfortunately, it is often neglected in the design of existing schedulers.

We use hierarchical GNNs to encode server architecture and network link information, and apply reinforcement learning to produce topology-aware scheduling policies, avoiding co-locating interference and mitigating cross-machine communication overhead at the same time. %Therefore, there is ample room for performance improvement through an interference-aware placement policy. While heuristics raise the generality issue, 
Applying one single RL scheduler to manage thousands of GPU servers may result in significant challenges as follow: (i) the high dimension of input state prevents the agent from learning valuable features; (ii) the very large action space takes long to be explored, and the model training is easy to fall into local optimum or become non-converging; (iii) the scheduler needs to perform a large number of inferences for all jobs submitted into the cluster, both in training and execution phases, leading to low training speed and extra scheduling delay. Instead, we design a multi-agent reinforcement learning framework, which allows multiple schedulers to cooperatively schedule DL jobs within limited neighborhood of the cluster.

\section{System Overview}
% 1 page
% 是否可以更简洁，把具体fat tree topology的定义放入下一section
\begin{figure*}[ht]
\centering
\includegraphics[width=1\textwidth]{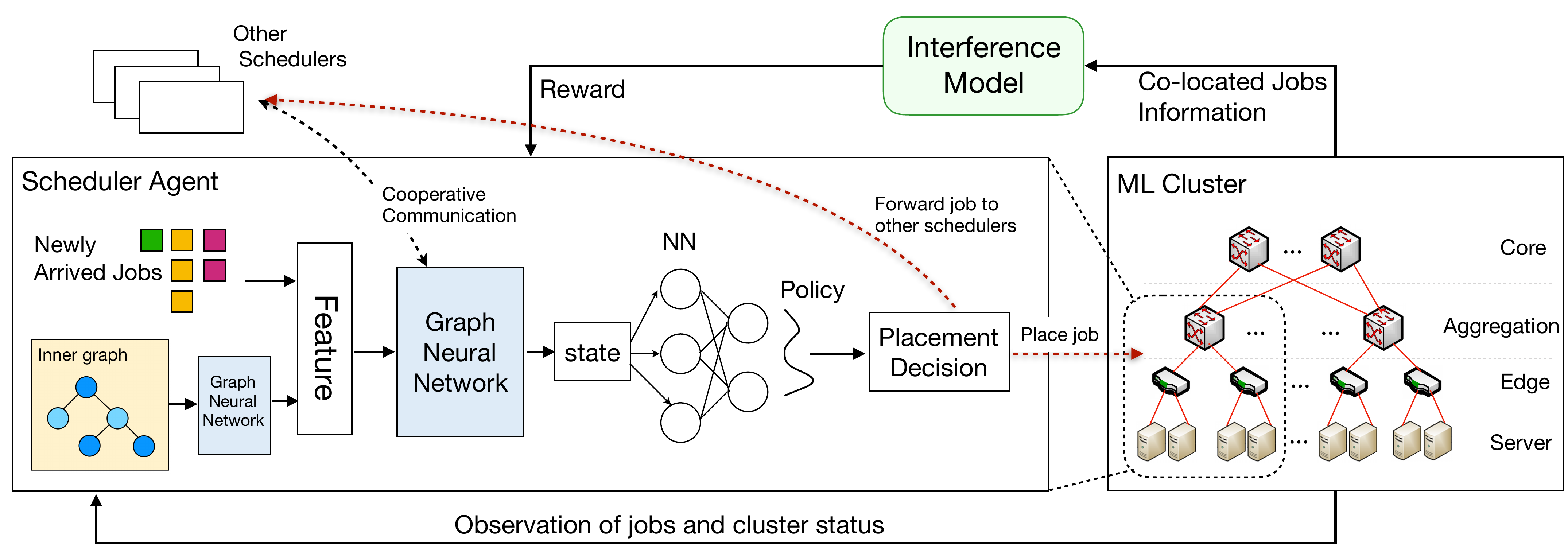}
\caption{Workflow of MARL schedulers}% and a customer of no one.}
\label{fig:overview}
\end{figure*}

\subsection{Multi-Scheduler Workflow}

We consider a large-scale machine learning cluster consisting of thousands of GPU servers. Data-parallel ML training jobs are submitted to the cluster by different users. Each job owner submits the following information: (i) the total number of workers (and parameter servers) it needs, which is determined according to the developer's experience on achieving better job performance (e.g., less convergence time, higher accuracy); (ii) The resource demand of each worker (and each PS), in terms of $L$ types of resources (e.g., CPU cores and GPU cards); (iii) The maximal number of training epochs required to complete the training process. Usually developers will set an upper bound on the number of epochs taken to achieve model convergence (indicated by the model loss or accuracy).

The cluster is managed by a number of schedulers (e.g., 20 at the scale of 2000 servers in our evaluation). Each job will be handled by one of these schedulers, decided based on the practice such as jobs submitted from the same group/team of the company are handled by one specified scheduler. 

% Communication, cooperation, importance
We divide the large ML cluster into partitions, e.g., each pod can be a partition in a datacenter using the Fat-tree~\cite{dc} topology. Each scheduler is responsible for placement of newly arrived jobs in each partition. If resources in the partition managed by a scheduler are not sufficient for hosting a job that the scheduler receives, the scheduler will forward the job to another scheduler. Schedulers can exchange their observations of server load status, link bandwidth usage and concurrent job placements. Once a job is placed, it will run to completion without resource preemption. This is currently the norm in production ML clusters~\cite{google-trace}, since any dynamic adjustment of resource allocation is hard to implement in practice.

\vspace{-3mm}
\subsection{Graph-based MARL}
% 本文框架的工作流程，obj, hirachical GNN，DRL, cluster trace, interference model used for reward （结合图）
% offline training, offline usage and update
% 参考 sigcomm, harmony

We propose a multi-agent reinforcement learning (MARL) framework (Fig.~\ref{fig:overview}) to make job placement decisions. The schedulers are the agents that use reinforcement learning (RL) to learn placement policies. Each scheduler (agent) takes as input the current \textit{state} of the cluster and outputs a scheduling \textit{action}. The overall objective is to minimize the average job completion time in the cluster.

At each scheduler, hierarchical GNNs are used for encoding the cluster state: the topology of the cluster partition that the scheduler is overseeing will be encoded by one GNN; this embedding and embeddings from other schedulers will be further encoded by another GNN. Encoding from the second GNN will be further processed by a policy network to learn placement decisions, towards larger training speed of jobs. 

%is  into consideration and encode it as an inner scheduler graph. At a high level, the scheduler targets to observe status of physical servers and multi-tier structure with links of different bandwidth, through a GNN. When placing workers/PSs of one job, it will pack GPUs as close as possible to obtain larger training speed, and avoid potential interference at the meantime (given current placement information on servers).

Exchanging information among schedulers through hierarchical GNNs enables strategical decisions on which other scheduler to forward a job's task(s) to, in case there is no available resource to host the job entirely at a scheduler.

The MARL framework is trained offline driven by a carefully designed DNN training interference model, and the learned policy will be executed online for job placement at the respective scheduler, upon new job arrivals. We detail our design of the MARL framework (Sec.~\ref{section:iv}) and the interference model (Sec.~\ref{section:v}) in the following sections.

% 详细说怎么offline，cluster trace/, 用reward引出interference model
% Offline training is indispensable for producing a good policy used online, especially when training NN through RL, which requires a large number of explorations; pure online training is time consuming and more likely to converge to a poor policy that is local optimal. We use various job sets containing different model types in Table~\ref{model-table} as the arriving jobs over time. Moreover, to make good placement decisions that mitigate the performance degradation due to resource contention, we design a fitting model to predict the performance slowdown (i.e., the degree of interference) when ML jobs with different resource specifications are co-located together. The interference model resolves the issue of inaccurate \textit{reward} (e.g., training speed) collected by schedulers in simulated offline experiments.

% We detail our design of DRL on each agent and the interference model in the following sections.

% 体现multi-agent, graph

\section{Graph-based MARL: Detailed Design}\label{section:iv}
% 2 page
In this section, we describe embedding generation of the hierarchical GNNs and present detailed MARL design that cooperatively generates job placement policies, with the objective of minimizing average job completion time.

\vspace{-3mm}
\subsection{Graph Embedding Generation}

A large-scale cluster is often structured using a multi-tier architecture (e.g., Fat-tree, BCube~\cite{dc}). We define the {\em inner graph} of a scheduler as the server/switch topology of the cluster partition managed by this scheduler, and organize all schedulers into an {\em inter scheduler graph}.

\begin{figure}
\centering
\includegraphics[width=0.35\textwidth]{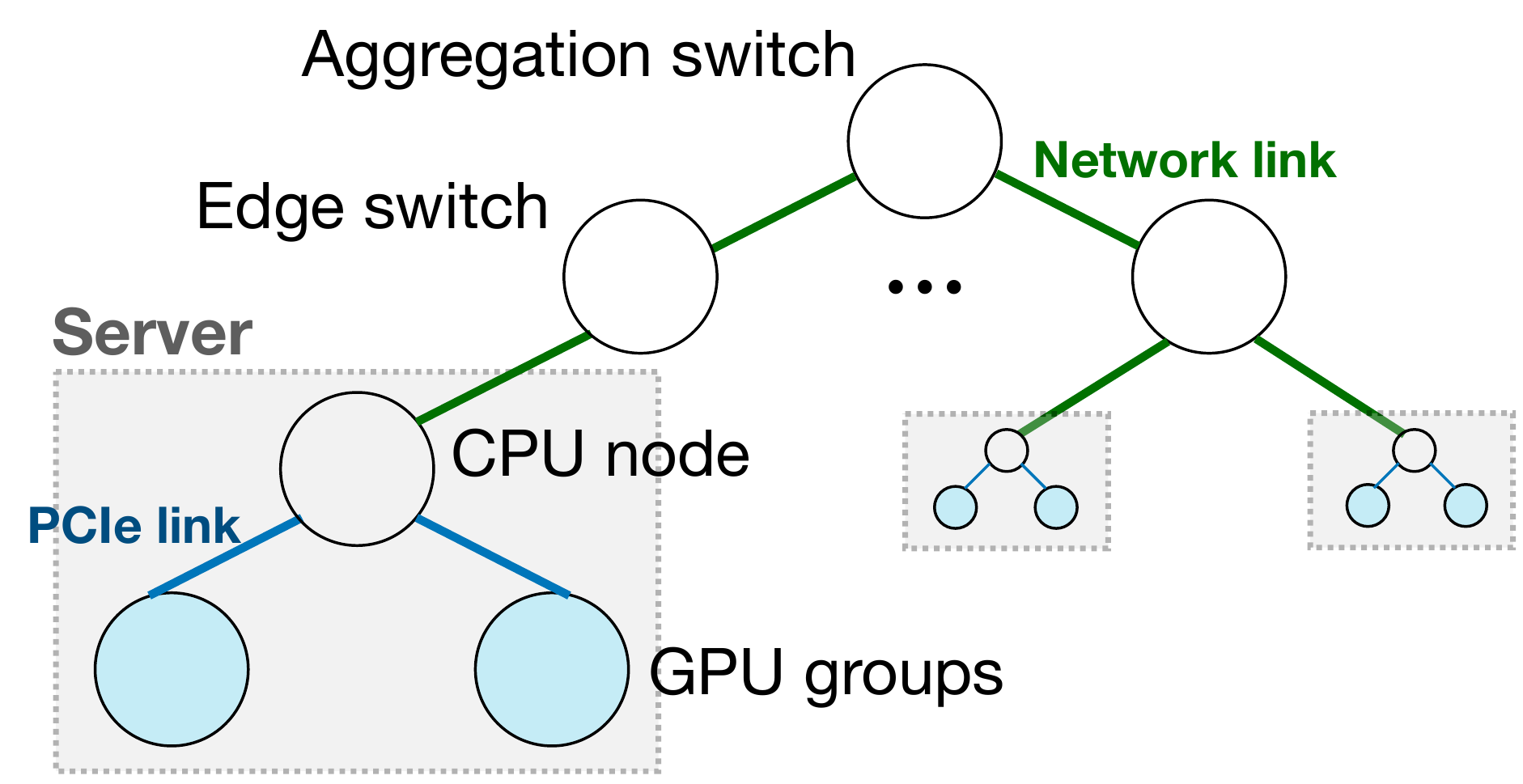}
\caption{Inner graph of one scheduler in a fat-tree cluster.}% and a customer of no one.}
\label{fig:graph}
\end{figure}

\vspace{1mm}
\noindent \textbf{Inner Graph of Each Scheduler.} We define the inner graph of a scheduler as $G_{inner}=(V_s, E_s)$, the architecture within each server is considered when construct this graph. For example, a GPU server may have two physical CPUs and eight GPUs; the GPUs are split into two groups and connected to two PCIe switches, respectively. We consider PCIe-only architecture in this work and model both physical CPUs and such GPU groups as nodes in the inner graph. Therefore, the node set $V_s$ of the inner graph contains physical CPUs on servers, GPU groups connecting to the same PCIe switch, as well as switches at the low tiers of the cluster architecture interconnecting those servers. The edges $E_s$ correspond to inter-connection between the nodes (i.e., network links between switches and servers, and PCIe connections between GPU groups and CPUs). An illustration of the inner graph is given in Fig.~\ref{fig:graph}.

For our inner graph embedding, we initialize node features, $h_u^0=(\vec{c}, \vec{d})$ for all nodes $u \in V_s$ that correspond to GPU groups. The node features are initialized as zero vectors for CPU or switch nodes. %$h_u^0$ will be the input to the first GNN.

% We adopt the following node features, $h_v^0=(\vec{c}, \vec{d})$ for all nodes $v \in V_s$ that correspond to GPU groups, for our inner graph embedding. The node features are initialized as zero vectors for CPU or switch nodes.

\vspace{1mm}
\noindent $\bullet$ $\vec{c}$ is an $L$-dimensional vector encoding available amount of each resource type on this node. Recall that $L$ is the number of resource types to compose a worker or a PS, e.g., GPU cards, CPU cores.

\vspace{1mm}
\noindent $\bullet$ $\vec{d}$ is a $2N$-dimensional vector representing current placement status of workers and PSs of concurrent jobs, where $N$ is the maximum number of concurrent jobs at the scheduler. The concurrent jobs include both newly arrived jobs and uncompleted jobs which were submitted earlier to the scheduler. The numbers of workers and PSs of job $n, \forall n \in [1,N]$ placed on the GPU group, are on the ${(2n-1)}^{th}$ and ${2n}^{th}$ position of $\vec{d}$, respectively (the PS number is zero if the job is using the AllReduce architecture). 

\vspace{1mm}
\noindent \textbf{Inter Scheduler Graph.} We define the inter-connection graph among schedulers as $G_{inter}=(V_c, E_c)$. Nodes set $V_c$ contains schedulers and switches at the top tier inter-connecting cluster partitions managed by the schedulers. Edge set $E_c$ includes the links connecting those switches and the respective partitions. Note that if there are more than one link connecting a top-tier switch to switch(es) in a partition, the link between the top-tier switch and the partition indicates an aggregate link with bandwidth as the sum of bandwidth on those physical links. 

Each scheduler node $v \in V_c$ in the graph has the following observation $o_v=(\mathbf{x}, \mathbf{r}, \mathcal{H}, \vec{p})$: 

\vspace{1mm}
\noindent $\bullet$ $\mathbf{x}$ is an $N \times Y$ binary matrix representing the model types trained by concurrent jobs, where $Y$ is the maximal number of model types that could be trained in the cluster. Each $Y$-dimensional vector $\vec{x}_n \in \textbf{x}, \forall n \in [1, N]$ is the one-hot encoding indicating the model type of job $n$. For example, in the case of 3 model types in the cluster (i.e., $Y=3$), we can use $x=\{[1,0,0], [0,1,0], [0,0,1]\}$ to indicate 3 concurrent jobs each training one model, respectively. Note that ML models with the same DNN architecture, training dataset and mini-batch size, but different learn rates or loss functions, still correspond to the same encoding.

% the same ML model, e.g., a DNN with the same architecture and mini-batch size but possibly different learning rates, corresponds to the same encoding.

\vspace{1mm}
\noindent $\bullet$ $\mathbf{r}$ is an $N \times 2(1+L)$ matrix encoding worker/PS resource demands of the concurrent jobs. In each vector $\vec{r}_n \in \textbf{r}, \forall n \in [1, N]$, the first element indicates the number of workers requested by job $n$, and the next $L$ values are the demand of each worker on $L$ types of resources; similarly, the remaining $(1+L)$ elements indicate the number of PS and the resource demand of each PS in job $n$. For example, $\vec{r}_n=[3, 3, 1, 2, 2, 0]$ implies that job $n$ has in total 3 workers, each consuming 3 CPU cores and 1 GPU, and 2 PSs, each using 2 CPU cores. % and no gpu

\vspace{1mm}
\noindent $\bullet$ $\mathcal{H}$ is an $M \times |h^0|$ matrix encoding embedding of GPU group nodes in $G_{inner}$ of the scheduler, where $M$ is the total number of GPU group nodes in the inner graph of the scheduler, and $|h^0|$ is the feature dimension of GPU group nodes. %This helps the scheduler be aware of inner topology and capture the status of concurrent jobs and GPU servers.

\vspace{1mm}
\noindent $\bullet$ $\Vec{p}$ is a vector of $(1+Y)+2(1+L)$ dimensions including information of jobs in the current inference (i.e., the jobs whose PSs/workers are to be placed). The first $(1+Y)$ values indicate the specific task (PS denoted by 1 and worker denoted by 0) and one-hot encoding of the model type; the next $2(1+L)$ elements denote the number of workers (PSes) and the resource demand of each worker (PS) in this job, similar as in the definition of $\mathbf{r}$.

% \vspace{1mm}
% The local observation $o_v$ is then encoded into a node feature vector $z^0_v$, by using an one-layer MLP for low-dimensional input.

 % information flow, how feature aggregated
%As shown in Fig.~\ref{fig:gcn}, 
\begin{figure}
\centering
\includegraphics[width=0.48\textwidth]{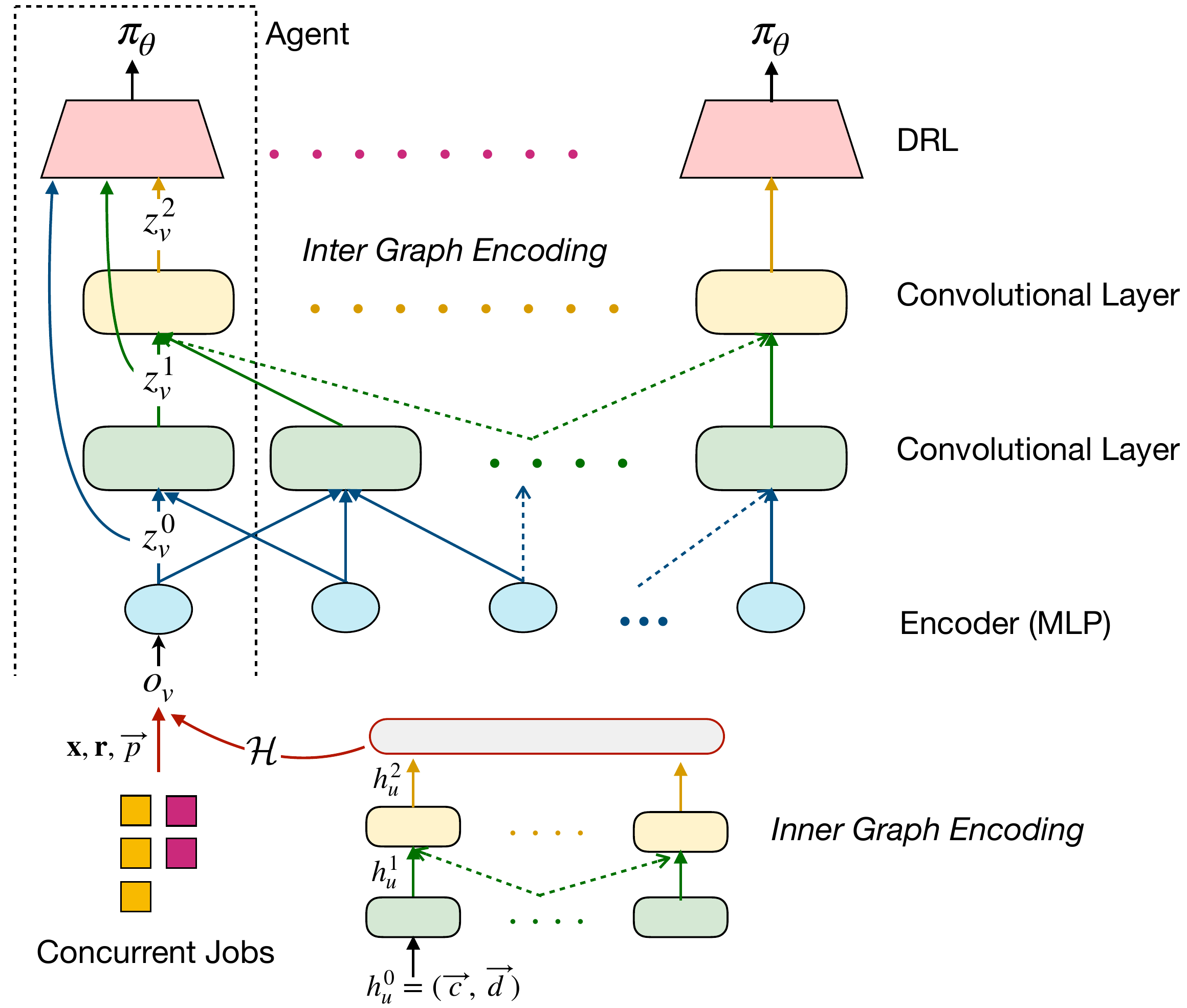}
\caption{The framework of multi-agent graph reinforcement learning. }
\label{fig:gcn}
\end{figure}

\vspace{1mm}
\noindent \textbf{GNN Embedding.} We use hierarchical GNNs with convolutional layers to aggregate features and generate embeddings (Fig.~\ref{fig:gcn}) for schedulers. At each scheduler (agent) $v \in V_c$, there are two GNNs, each for encoding the inner graph of the scheduler and inter scheduler graph, respectively. The input to the first convolutional layer of the first GNN is $h_u^0, \forall u \in V_s$. After a few convolutional layers, the produced embeddings $\mathcal{H}$ will be combined with $(\mathbf{x}, \mathbf{r}, \Vec{p})$ to form the scheduler's observation $o_v$, which will then be fed into an MLP encoder (for dimension reduction) to generate the scheduler's node feature vector $z_v^0$. $z_v^0, \forall v \in V_c$, is the input to the first convolutional layer of the second GNN. Each scheduler further aggregates other schedulers' feature vectors through multiple convolutional layers and generates inter scheduler graph embedding.
% (4/2 as in our evaluation)

% We use hierarchical GNNs with convolutional layers to aggregate features and generate embeddings (Fig.~\ref{fig:gcn}). At each scheduler (agent), there are two GNNs, each for encoding the inner graph of the scheduler and inter scheduler graph, respectively. The input to the first convolutional layer of the first GNN is $h_v^0, \forall v \in V_s$. After a few convolutional layers (4 as in our evaluation), an embedding $\mathcal{H}$ of the inner graph of the scheduler is produced. The embedding is further combined with the scheduler's observation $o_v$ to generate a node feature vector $z_v$. The scheduler further aggregates other schedulers' node feature vectors, and feeds all into the second GNN for inter scheduler graph encoding. 

%The embedding generation process for $G_{inner}$ and $G_{inter}$ takes features for all nodes $h_v^0, \forall v \in V_s$ and $z_v^0, \forall v \in V_c$ as input, respectively. 
At each convolutional layer $k$ in each GNN, a node $u$ ($v$) aggregates representations of the nodes in its immediate neighboring set $\mathcal{N}_u$ ($\mathcal{N}_v$) from the last layer $k-1$:

\begin{center}
    $h^{k}_{\mathcal{N}_u}=g(\{h^{k-1}_w, \forall w \in \mathcal{N}_u\}), \forall u \in V_s$\\[2mm]
    $z^{k}_{\mathcal{N}_v}=g(\{h^{k-1}_w, \forall w \in \mathcal{N}_v\}), \forall v \in V_c$
\end{center} 
where $g$ is the node feature aggregator, e.g., a $mean$ aggregator naturally for graph convolution.

%Most distributed training jobs are communication-bound, which means the communication cost between PS and worker (or among workers for all-reduce) will influence the training performance most. However, there are lots of cases that one job's workers and PSs cannot be packed locally within one server, e.g., server unavailable due to resource fragment or potential interference due to resource contention. Hence, when the scheduler are placing the worker/PS of one job, the link bandwidth should also be considered since it determined the time cost on exchanging gradients and updated parameters.

% system example, why ecc
To better capture the edge information in both inner graph of the scheduler and inter scheduler graph, we adopt Edge Condition Convolution (ECC)~\cite{ecc} as the node feature aggregator $g$ to achieve $weighted \text{ } mean$ aggregation. For the example of $G_{inner}$ encoding, to obtain neighboring feature vector $h^k_{\mathcal{N}_u}$ of node $u \in V_s$ at convolutional layer $k$, we aggregate neighbors' embeddings at layer $k-1$ as follows:

\begin{equation}
\begin{aligned}
    h^k_{\mathcal{N}_u}&=\frac{1}{|\mathcal{N}_u|}\sum_{w \in \mathcal{N}_u} F^k\big(E(u, w);\phi^k\big)h^{k-1}_w + b^k\nonumber\\
    &= \frac{1}{|\mathcal{N}_u|} \sum_{w \in \mathcal{N}_u} \Theta^k_{uw} h_w^{k-1} + b^k
\end{aligned}
\end{equation}

\noindent where $F^k$ is an one-layer linear NN with learnable parameters $\phi^k$, which transfers the low-dimension edge vector $E(u,w)$ into the aggregation weight $\Theta^k_{uw}$ between neighboring node $u$ and $w$; $b^k$ is the bias value. The edge vector contains information about bandwidth, load status and the tier where the link is located in the data center.

Then the node’s representation at layer $k-1$ is concatenated (denoted by the $[\cdot]$ operator) with the aggregated neighboring vector, and this concatenated vector is fed through a fully connected layer with nonlinear activation function $\sigma$ and learnable weight matrix $\textbf{W}^k$, which transforms the representations to be used at the layer $k$. Hence, for both GNNs, the update rule of node feature can be presented as:
\begin{center}
    $h^k_u=\sigma_h(\textbf{W}^k_h \cdot [h_u^{k-1}, h_{\mathcal{N}_u}^k]), \forall u \in V_s$\\[2mm]
    $z^k_v=\sigma_z(\textbf{W}^k_z \cdot [z_v^{k-1}, z_{\mathcal{N}_v}^k]), \forall v \in V_c$
\end{center}
The final representations output at layer $K$ will be the embedded features of nodes in the respective graph. The embedded feature after inter scheduler graph aggregation will be the input state to the scheduler's DRL module, as illustrated in Fig.~\ref{fig:gcn}.

By stacking more convolutional layers in the GNN that encoding the inter scheduler graph, the receptive field of a scheduler grows, i.e., it could gather more information of other schedulers for better learning which server is suitable to undertake the workload and which communication routes have smaller latency/larger bandwidth for distributed training. 

\vspace{-3mm}
\subsection{DRL Formulation}

We next detail the DRL module at each scheduler, which is used to generate specific job placement policy after the encoding of hierarchical GNNs.

\noindent \textbf{State Space.} %After the graph embedding generation process, we further define
The input state to scheduler $v$'s %\in V_c$'s 
DRL model is $s_v=(z_v^0, z_v^1, ..., z_v^K), \forall v \in V_c$. As inspired by DenseNet~\cite{example}, we concatenate the features of all layers of the GNN encoding the inter scheduler graph as input into the policy network, in order to assemble and reuse the observation representations and features from different receptive fields. 
%This integration contributions to the strategy that takes the cooperation at different scopes into consideration.

\vspace{1mm}
\noindent \textbf{Action Space.} %After collecting $s_v$, 
Each scheduler $v$ produces an action $a_v$ based on policy $\pi_\theta^v(a_v|s_v)$, which is a probability distribution over the action space. The policy is produced by an NN with $\theta$ as the set of parameters. The action space at each scheduler contains $M_v+|V_c\setminus\mathbb{S}|$ actions, where $M_v$ is the number of GPU groups in scheduler $v$'s inner graph, $\mathbb{S}$ is the set of switches in the cluster, and $|V_c\setminus\mathbb{S}|$ indicates the number of schedulers in the cluster. Each action indicates a placement choice of a worker or PS of the job being considered, i.e., on one of the $M_v$ GPU groups managed by this scheduler, or being forwarded to one of the $|V_c\setminus\mathbb{S}|$ other schedulers for placement. We perform multiple inferences to produce decisions for all workers/PSs sequentially, in order to produce complete placements of a given job.

\vspace{-3mm}
\subsection{Model Training}
% 定义loss，

Our GNNs and policy NN are trained together using offline simulation. Given job arrival traces, we divide time into equal scheduling intervals, and train our model using jobs arrived in each scheduling interval in a batch processing manner.

\vspace{1mm}
\noindent \textbf{Reward.} Each scheduler targets average job completion time minimization in training its policy NN. We use the normalized training speed of each job (i.e., the number of trained epochs in a scheduling interval divided by the maximum epochs needed for training the job) as the reward in creating a training sample. The reward implies job training progress: the more progress achieved in one interval, the fewer intervals a job takes to complete training. Hence, for each scheduler, maximizing cumulative reward is equivalent to minimizing average job completion time of concurrent jobs.

\vspace{1mm}
\noindent \textbf{Sample.}
At each scheduler, we produce one training sample per job it scheduled at the end of each scheduling interval, in the form of $<s,a,r,s'>$: $a$ is the placement decision produced for the respective worker/PS, $s$ and $s'$ are the input states before and after the placement action $a$; $r$ is the reward computed in the current interval (workers/PSs in the same job have the same reward). 
% The goal of each scheduler is to maximize the expected cumulative discounted reward $E[\sum_{t \geq 0} \gamma^t r^t]$, where $\gamma \in [0, 1]$ is the discount factor and $t$ represents the number of inferences made.

\vspace{1mm}
\noindent \textbf{Training algorithm.}
The policy NN on each scheduler aims to maximize the expected cumulative discounted reward $\mathcal{J}(\theta)=E[\sum_{t \geq 0} \gamma^t r_t]$, where $\gamma \in [0, 1]$ is the discount factor and $t$ represents the number of inferences made. The policy gradient used for NN update can be computed as:
\begin{center}
    $\nabla_\theta \mathcal{J}(\theta)=E_{\pi_\theta}[\sum\limits_t \nabla_\theta log(\pi_\theta(a_t|s_t))Q^{\pi_\theta}(s_t, a_t)]$
\end{center}
where the $Q$ value indicates the \textit{quality} of action $a$ taken in given state $s$ following the policy $\pi_\theta$, usually calculated as expected cumulative discounted reward to obtain after selecting $a$ under $s$ following $\pi_\theta$, i.e., $Q^{\pi_\theta}(s_t, a_t)=\sum_{t'\geq t}\gamma^{t'-t}r_{t'}$.
% Q 是qualtiy，会有variance，用下面那个 adv代替 Q

To prevent high variance in Q values and ensure expedited convergence of the policy network, we adopt \textit{actor-critic} algorithm~\cite{marl}, which introduces an advantage function to replace $Q$ value for policy gradient calculation, i.e., $\delta = r + \gamma V^{\pi_\theta}(s'; \omega) - V^{\pi_\theta}(s; \omega)$, where $V^{\pi_\theta}(s; \omega)$ is calculated as the expected cumulative reward following the policy $\pi_\theta$ from state $s$, over all possible actions; we use a value network (the \textit{critic}) with $\omega$ as the set of parameters to estimate $V^{\pi_\theta}(s;\omega)$. Specifically, the \textit{actor} is a policy network with input $s$, and output $\pi_\theta(s)$; the \textit{critic} has input $s$, but the output layer is a linear neuron without any activation function. The loss functions of \textit{actor} and \textit{critic} on each scheduler are defined as:
\begin{align*}
    &\mathcal{L}_{actor}(\theta) = -\delta log\pi_\theta(s); \quad \mathcal{L}_{critic}(\omega) = \delta ^2
\end{align*}

% To allow effective exploration in training, we use an $\epsilon-$greedy approach~\cite{marl} to adequately explore the action space: with each inference, an agent has probability $1-\epsilon$ to choose the action produced by the current policy $\pi_\theta(s)$, and has probability $\epsilon$ to randomly choose between multi-resource bin packing and load balancing policies, and place the job according to chosen policy.

In our offline training of the policy network, we quantify the impact of interference among co-located jobs on a server with an interference model that we build. We detail the interference model in the following section.

% ================================================= 

\section{Interference Model}\label{section:v}

To make better placement decisions that mitigate performance degradation due to resource contention, we carefully design an interference model to predict the training slowdown when ML jobs with different resource specifications are co-located on the same server. In the offline training phase, the interference model is used to produce more accurate training speeds given any placement of co-located jobs (i.e., even unseen among historical samples).
% (that requires a large number of reward samples as feedback for a chosen scheduling policy)

\vspace{-3mm}
\subsection{Metrics}
During one mini-batch of ML model training, disk I/O is firstly envoked to read training data from the disk to the host memory; training samples after being pre-processed in CPU will be packed into tensors and transferred to GPU memory through PCIe link; calculated gradients are then pushed and pulled between CPU and GPU, traversing PCIe. %For most DL models, mini-batch cycles are generally quite small, and mini-batches apply mostly similar computations that can be utilized to profile the job performance.

We use PCIe bandwidth and CPU utilization as profiling metrics to characterize resource contention because they represent more frequently exploited resources during training. 

% but in most DL frameworks~\cite{tensorflow}, \textit{DataLoader} is used as a tool to fetch training data. With \textit{DataLoader}, instead of requesting disk I/O in every mini-batch,

Disk I/O is also shared among co-located jobs, but developers usually use \textit{DataLoader}~\cite{tensorflow} to periodically fetch training data for more than one mini-batch to avoid frequent I/O requests. To verify this, we run one job training ResNet50 with ImageNet in a server (with one Intel E5-1660 CPU, two 1080Ti GPUs, and 48GB RAM) and record its average data loading time per mini-batch in two cases: (1) \textit{DataLoader}  requests Disk I/O in every mini-batch; (2) \textit{DataLoader} requests Disk I/O once every 20 mini-batches. We can observe from the results in Table~\ref{dataloder} that frequent I/O requests lead to 
significant slowdown of data loading, when we increase the number of co-located jobs. However, the I/O contention could be eliminated by a periodical setting of \textit{DataLoader}, as different jobs may fetch data at different times. Therefore, we don't consider disk I/O as one metric in our interference model.

% The process repeats with a few short periods of varying resource demand, e.g., testing phase, saving checkpoint. 

\vspace{-3mm}
\subsection{Model}

We define the training slowdown of a ML training job as $\frac{T_{real}-T_{alone}}{T_{alone}}$, where $T_{real}$ is a job $J$’s completion time when co-located with other jobs $\tilde J$ on the same server server and $T_{alone}$ is the job's running time by running it alone on the server. %We use $J$ to denote the job that we target to predict its slowdown, and use $\tilde J$ to indicate those jobs that co-locate with $J$ in the same server.
%According to the PCIe-only server architecture, 
We use $\mathcal{G}_J^{same} \subseteq \tilde J $ to indicate jobs that run on the same GPU group as job $J$, i.e., connected to the same CPU in a PCIe-only server architecture; we use $\mathcal{G}_J^{diff} \subseteq \tilde J $ to indicate jobs that belong to a different GPU group from job $J$, i.e., connected to a different CPU in the server. Jobs within the same group use the same PCIe bus to communicate with the same CPU. To predict the slowdown of job $J$ when co-located with $\tilde J$, denoted as $S(J, \tilde J)$, we consider the interference due to CPU sharing and PCIe sharing separately, as CPU processing and PCIe bus transmission can be treated as isolated phases in most DL training cases. 

\begin{table}
\small
\centering
\caption{Average Data Load Time of training ResNet50 on ImageNet.}\label{dataloder}
\begin{tabular}{|c|c|c|}
\hline
\# of co-located jobs   & case (1)  & case (2)  \\ \hline
0   & 0.007 s  & 0.004 s \\
1   & 0.088 s  & 0.004 s \\
2   & 0.236 s  & 0.007 s \\
3   & 0.339 s  & 0.009 s \\
4   & 0.420 s  & 0.012 s \\ \hline
\end{tabular}
\end{table}

\vspace{1mm}
\noindent \textbf{Slowdown due to CPU Interference.} We use an exponential model to predict job slowdown due to pure CPU interference with co-located jobs. %For an application, 
Typically there is no significant performance degradation until the overall CPU utilization in a server approaches the number of physical cores. The reason is that in the architecture of a physical CPU, L1/L2 cache is isolated for each physical core~\cite{gexin}. The contention of cache occurs when demands exceed the number of physical cores, which is common since one physical core is typically virtualized into two logical cores for DL developers (through Hyper Thread)~\cite{topology17}. The slowdown of job $J$ due to CPU interference with co-located jobs $\tilde J$ is modeled as:

\begin{center}
    $S_{cpu}(J, \tilde J) = \alpha_1 exp\big(\alpha_2 U_c(\tilde J) + \alpha_3 \mathcal{C}_J\big) + \lambda_1$
\end{center}
where $\mathcal{C}_J$ is the CPU utilization of training %the same model as 
job $J$ (profiled by NVIDIA profiler tools~\cite{profiler} when running alone), and $U_c(\tilde J)$ indicates total CPU utilization of the interfering jobs:
\begin{equation*}
    U_c(\tilde J) = \sum_{j\in \mathcal{G}_J^{same}} \mathcal{C}_j + \big(\sum_{j \in \mathcal{G}_J^{diff}} \mathcal{C}_j - n_{core} \big)_{+}
\end{equation*}
where $n_{core}$ is the number of cores in one CPU. The rationale is that a running job will firstly utilize the CPU with affinity, if the utilization of affinity CPU reaches capacity. The job will then utilize the other CPU, which causes contention with existing jobs running on it.
% The performance of a job will also be impacted by jobs running on the other CPU, when that CPU's capacity is reached. %\cwu{revise the sentence as the idea is not clear}
$\alpha_1, \alpha_2, \alpha_3$ and $\lambda_1$ represent coefficients and constant to be fit in the model. 

% Jobs running on the other CPU will still degrade performance of jobs  
% We consider that jobs attached to a different CPU could also degrade performance when CPU capacity is reached

%Jobs with different model types have corresponding resource sensitivity. Usually the job with larger inherent resource utilization is more sensitive to the contention from interfering jobs. Hence, in the formulation of $\hat S_{cpu}$, we adopt $Cpu(j)$ as part of the exponent to reflect the sensitivity to CPU interference.

To fit the model, we vary the types of jobs (i.e., training different ML models), and use a CPU workload generator conducting arithmetic operations to generate different levels of interfering CPU usage, for the purpose of collecting samples for model fitting.

\vspace{1mm}
\noindent \textbf{Slowdown due to PCIe Interference.} In a typical server architecture, CPUs are connected using inter-socket (e.g., QPI), which can have a bandwidth higher than 300 Gbps~\cite{topology17}. 
The inter-connection link between PCIe switch and CPU is more likely to become the bottleneck, since it is frequently used for transmitting gradients from a group of GPU devices. Hence, we consider job $J$ competes with jobs exploiting the same CPU (i.e., $\mathcal{G}_J^{same}$) for PCIe bandwidth, and build a linear model to predict job $J$'s slowdown due to PCIe contention:
\begin{center}
    $S_p(J, \tilde J)=\beta_1 U_p(\tilde J) + \beta_2 \mathcal{P}_J + \lambda_2$
\end{center}
where $\mathcal{P}_J$ is the PCIe usage of the job when running alone, and we use NVIDIA profiler tools to profile the usage data. $U_p(\tilde J)$ is the total PCIe usage of interfering jobs:
\begin{equation*}
    U_p(\tilde J) = \sum_{j \in \mathcal{G}_J^{same}} \mathcal{P}_j
\end{equation*}
$\beta_1,\beta_2$ and $\lambda_2$ are coefficients and constant in the model to be fit. We vary the types of training jobs and their placements on servers to collect samples for model fitting.

\vspace{1mm}
\noindent {\bf Overall interference model.}
We use $S(J,\tilde J)=S_{cpu} + S_p$ to predict training slowdown of job $J$, when co-located with job set $\tilde J$. The interference model is applied in our trace-driven evaluation to generate practical reward samples for learning.

\section{Trace-driven Simulation}

% \begin{figure*}[ht]
% \centering
% \includegraphics[width=0.85\textwidth]{figures/exp-job-width.pdf}
% \caption{Performance comparison under different job arrival patterns.}% and a customer of no one.}
% \label{fig:exp-job}
% \end{figure*}

% \begin{figure*}[ht]
% \centering
% \includegraphics[width=0.85\textwidth]{figures/exp-server-width.pdf}
% \caption{Performance comparison under different server configurations.}% and a customer of no one.}
% \label{fig:exp-server}
% \end{figure*}

% 2 page
%In this section, we will compare our framework with other scheduling schemes under different configurations, in terms of average job completion. We also verify that using MARL could achieve better convergence and job performance, as compared to applying traditional RL. Prediction error of the interference model is also analysed.

\subsection{Evaluation Methodology}
\noindent \textbf{Trace Workload.} By default, offline training workloads arrive at each scheduler following the job arrival pattern (i.e., the number of arrived jobs per scheduling interval) extracted from Google cluster traces~\cite{google-trace}, with down-scaled arrival rates. Upon a job arrival, we randomly select one from the 8 model types listed in Table~\ref{ml-table} as the training model. The number of workers/PSs required by one job is randomly set in $[1,4]$. Other job information, i.e., total number of epochs, resource demands, model size and mini-batch size, is set based on representative implementation of the models~\cite{example}. Scheduling interval is 30 minutes, and schedulers need to produce placement policies for a batch of newly arrived jobs. For each scheduler, the total job set is split into training dataset (90\%) and testing dataset (10\%). The training dataset is used for policy network training as described above, and the testing dataset is used for online inferences (i.e., the placement of a job is determined upon its arrival). We use the average job completion time (JCT) as the performance metric, and show the evaluation results on the testing dataset.
% (around 1000 jobs)

\noindent \textbf{Simulator.} We conduct trace-driven simulation to evaluate our framework with representative scheduling schemes under different configurations. The evaluation is run on a server with one Intel E5-1660 CPU, two GTX 1080Ti GPUs, 48GB RAM, and one MCX413A-GCAT 50GbE NIC. We simulate the data center topology and server architecture of a large-scale cluster, DL workloads will arrive at each scheduler following certain pattern for offline training. %Details are presented as follows.
For worker computation time, we first calculate the ideal time as the time of training with one worker on the whole dataset (profiled in advance) divided by the number of assigned workers, because in data-parallel each worker is trained on a partition of dataset. Then we multiply the ideal time by the slow-down factor computed from our interference model. 
Per worker pair communication time is calculated by dividing the transmitted gradient size by the bottleneck bandwidth in the network. Since we consider synchronize training, we set one iteration time as the sum of straggler's computation time and communication time.

% We conduct our evaluation on a server with one Intel E5-1660 CPU, two GTX 1080Ti GPUs, 48GB RAM, and one MCX413A-GCAT 50GbE NIC. %To expedite the training of multiple agents, we equally allocate agent models onto two GPUs.

\vspace{1mm} 
\noindent \textbf{DNN Implementation.} For offline training, we implement the DRL model in every scheduler using PyTorch libraries. The policy network (\textit{Actor}) and the value network (\textit{Critic}) have one hidden layer with 128 neurons, respectively. The GNN for encoding the inter scheduler graph has 2 convolutional layers; the GNN that encodes the inner graph of each scheduler uses 4 convolutional layers. While stacking more layers will increase the reception field of each graph node in GNNs, it may also bring about over-smooth feature problem when the GNN depth is large. We adopt \textit{ReLU} activation function for the hidden layer and \textit{Adam} Optimizer with learning rate $10^{-5}$, and set reward discount factor to $\gamma=0.9$. 

\vspace{1mm}
\noindent \textbf{Cluster Topology.} By default, our cluster topology is a $k$-ary fat-tree, containing switches with $k$ ports. Specifically, we consider a cluster topology with the common setting~\cite{fat-tree} $k=20$, containing 20 pods (one pod is managed by one scheduler) and $(\frac{k}{2})^2=100$ core switches. Each pod has $k/2=10$ edge switches, $k/2=10$ aggregation switches and $(\frac{k}{2})^2=100$ physical GPU servers. Link bandwidths in different tiers are 10Gbps, 20Gbps and 40Gbps respectively. To simplify the topology (in terms of both density and layers), we fuse aggregation switch nodes into one group node, since they are fully connected with switches in other tiers. Totally 2000 servers are simulated, which is at the scale of ML clusters operated by large IT companies to support their services.

\vspace{1mm}
\noindent \textbf{Server Architecture.} By default, servers in the cluster have a system architecture similar to the IBM Power8 system~\cite{topology17}, consisting of four GPUs and two CPUs with two GPUs per CPU. Each CPU has 8 cores and communicates with attached GPUs through PCIe-v3 with bandwidth 128Gbps. CPUs are connected using QPI, which supports quick path inter socket with bandwidth 300Gbps.

% At each scheduler, the total job set (around 1000 jobs) is split into training dataset (90\%) and testing dataset (10\%). The training dataset is used for policy network training and the testing dataset is used for online inferences (i.e., the placement of a job is determined upon its arrival). We use the average job completion time (JCT) as the performance metric, and evaluate it on the testing dataset.

\vspace{1mm}
\noindent \textbf{Baselines.} We compare our scheduling framework with three respective scheduling schemes:

- Tetris~\cite{tetris}: it uses multi-resource bin-packing to place workers/PSs, which follows the locality principle and avoids resource fragment. Jobs are maximally packed onto servers to avoid potential cross-machine communication and minimize resource fragmentation.

- Load Balancing (LB): it is adopted by some practical schedulers, e.g., Mesos~\cite{mesos}, Kubernetes~\cite{eks}. In each server, the usage of different types of resources is normalized and summed up as the server load. We always assign a worker or a PS to the server with the least load.

- Least Interference First (LIF): we use our designed interference model to score the interference level on each server, which is calculated as the execution slowdown of all tasks on a server. In this heuristic-driven scheme, a worker/PS will be placed onto the server with the least interference.

- DeepSys~\cite{deepsys}: it builds a DNN model to generate job speed prediction. Input of the speed model includes job model type, number of PS/worker, placement of each worker and the parameter size of all the existing workers on the server where the job runs. The output is a normalized speed value. We use historical placement information collected to train this model. Whenever a job arrives, multiple inferences are conducted to search for the placement achieving the largest job speed.

- Attentive Scheduling (Atte): we adopt the placement policy in SCARL~\cite{scarl}, where upon a job arrival, for each worker, the scheduler calculates an importance score for each server using the attention mechanism~\cite{attention}, and chooses the one with the largest score to place the worker.

%which determines the job processing order, which is not our concern, we only use the placement policy it proposed.

Our framework has two variants:

- Server-level: when construct the inner graph $G_{inner}$ of scheduler, we use servers and switches as nodes instead of CPU and GPU groups. When a worker/PS is placed onto a server, we will further randomly assign it to one GPU group.

% we use servers and switches as nodes in the inner graph of each scheduler $G_{inner}$, but not CPU and GPU groups in each server. When a worker/PS is placed onto a server, we will further randomly assign it to one GPU group.

- Device-level: inner graph $G_{inner}$ is defined as in Sec.~\ref{section:iv}-A.
Hence the generated scheduling policy not only decides a server, but also determines the specific CPU socket (i.e., attached GPU group) to place each worker/PS.

\vspace{-3mm}
\subsection{Impact of Job Arrival Patterns}
\begin{figure}[t]
\centering
\includegraphics[width=0.45\textwidth]{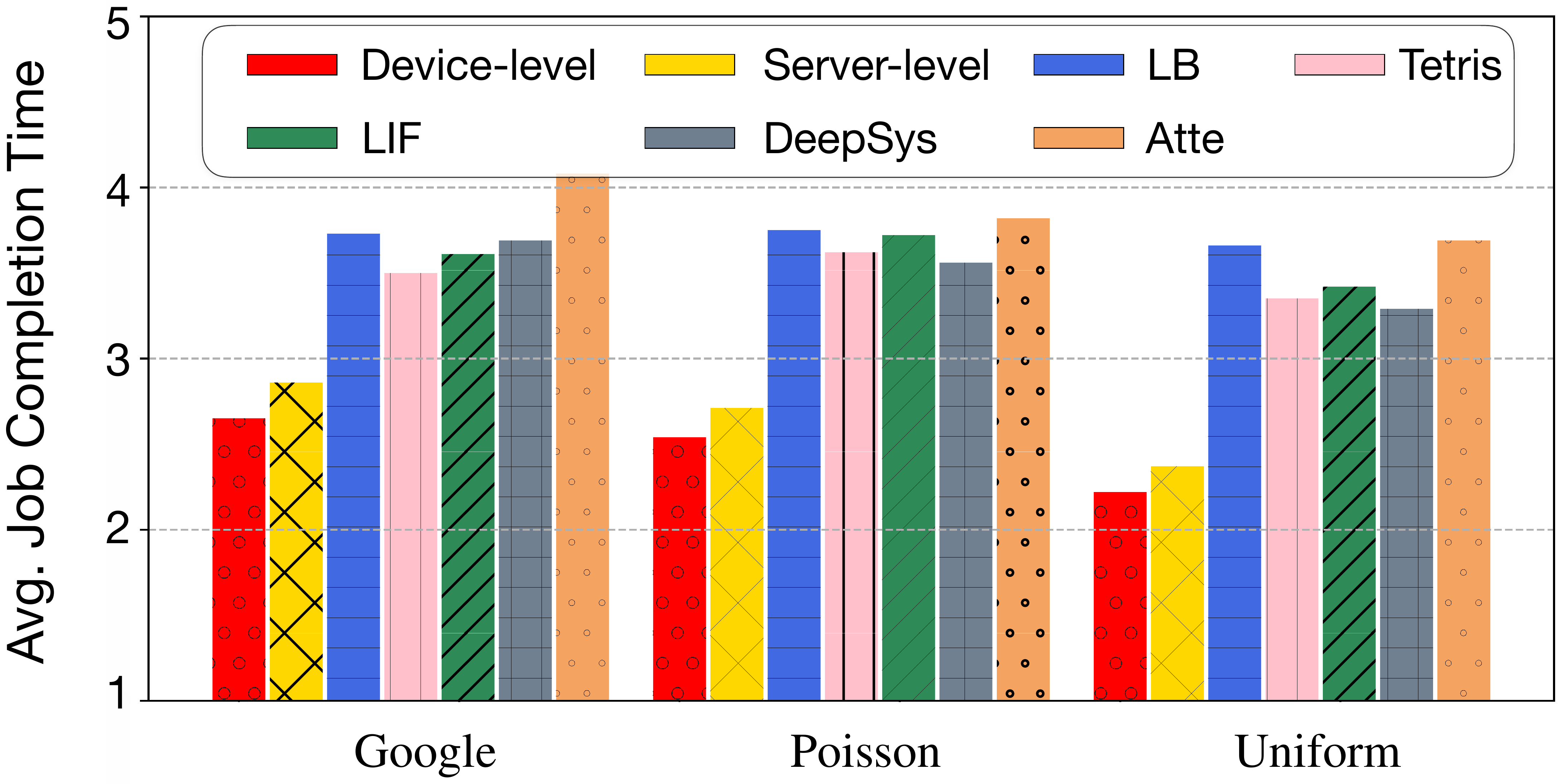}
\caption{Performance under different job arrival patterns.}
\label{fig:exp-job}
\end{figure}

% choose uniform (round robin), Poisson, google trace
Fig.~\ref{fig:exp-job} compares the performance of our scheduling framework with baselines under different jobs patterns, where the job completion time is in terms of  the number of scheduling intervals. With the uniform pattern, jobs are submitted to each scheduler in a uniformly random manner of 15 jobs per scheduling interval; the Poisson pattern corresponds to a Poisson process with an arrival rate of 15 jobs per interval. %; Google trace is the default workload setting. Server architecture and cluster topology are set as default.

We observe at least 24.3\% improvement in terms of average JCT as compared to all baselines.
% We observe at least 24.3\% improvement in terms of average JCT as compared to Tetris, 28.9\% as compared to Load balancing, and 26.6\% as compared to LIF. 
The performance of Tetris is worse, since the bin packing strategy targets to increase resource utilization but fails to handle the inter-job interference. Load Balancing focuses on reducing potential interference on each server, but neglects the communication overhead that results. Especially in a large-scale cluster, poor placements of worker/PS, e.g., multi-hop routing required, congested links traversed, could lead to significant slowdown for a synchronous training job. Although Least Interference First scheme achieves better interference avoidance using the explicit interference model, the same problem exists as load balancing. 
The performance of DeepSys heavily relies on the predicting accuracy of the speed model. However, it is difficult for such a model to aggregate all experiences in today's clusters of thousands of servers for the best accuracy. Atte is facing the same scalability issue, whose attention mechanism may fail to boost important parts of the features. Further, both ML-based methods are interference oblivious.

Our scheduling policy learns a good trade-off between load balancing and locality preserving. It mitigates computation interference and reduces transmission cost at the same time. Further, the performance of our device-level framework is better than the server-level variation. The reason lies in that the placement within one server also influences inter job interference; jobs co-located under the same CPU experience more serious interference than under different CPUs.

\begin{figure}[t]
\centering
\includegraphics[width=0.45\textwidth]{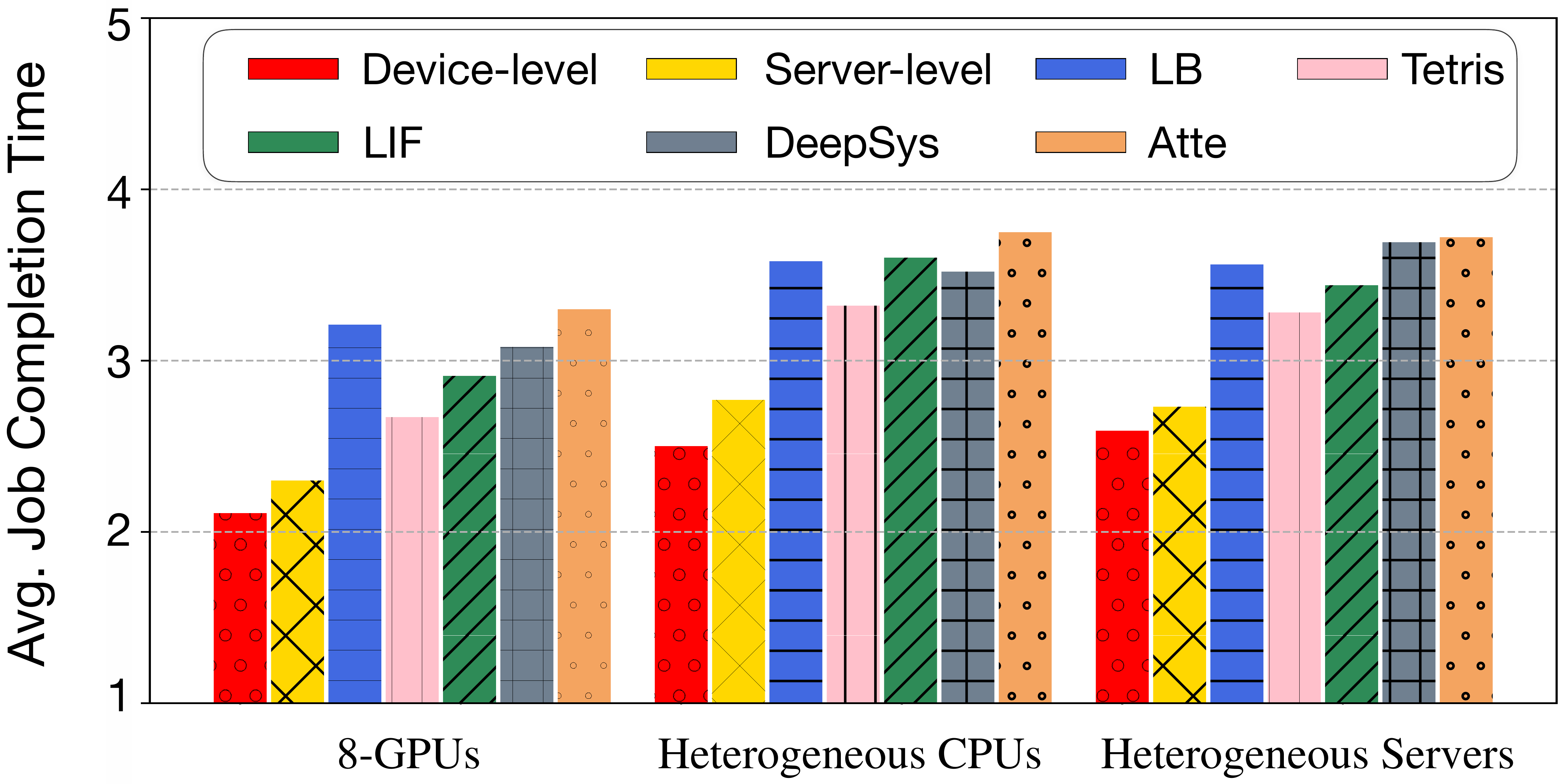}
\caption{Performance under different server configurations.}
\label{fig:exp-server}
\end{figure}

\vspace{-3mm}
\subsection{Performance in case of Heterogeneous Servers}

%Servers adopt architecture consisting of 4 GPUs and 2 CPUs by default. 
To evaluate the generality of our proposed framework under different server configurations, we vary server architectures as follows: (i) servers adopt architecture similar to a DGX-1 system~\cite{topology17}, consisting of 8 GPUs and 2 CPUs with 4 GPUs per one. Each CPU has 16 cores and communicates with attached GPUs through PCIe switch; (ii) servers have heterogeneous CPU configurations, with each server having 2 CPUs in total, one (16 cores) attached with 4 GPUs and the other (8 cores) with 2 GPUs; (iii) heterogeneous server configurations, where for the partition managed by each scheduler, 20\% of the servers each have 2 GPUs and 1 CPU (8 cores), 40\% adopt the 4-GPU architecture as our default, the rest 40\% adopt the 8-GPU architecture. %Cluster topology and workload are set as default.

In Fig.~\ref{fig:exp-server}, %compares the scheduling performance under the two cases,
we can observe that our learnt scheduling policy can achieve at least 22\% improvement as compared to baselines in all server architecture settings. With the help of GNNs, our schedulers can encode inner graphs to capture detailed server architecture and placement status, enabling generality over various heterogeneous server configurations.

\vspace{-3mm}
\subsection{Adaptability to Cluster Topology}

% \begin{figure}[t]
% \centering
% \includegraphics[width=0.48\textwidth]{figures/exp-cluster.pdf}
% \caption{Performance comparison under different cluster topologies.}% and a customer of no one.}
% \label{fig:exp-cluster}
% \end{figure}

To investigate the adaptability of our framework to different cluster topologies, we evaluate it under different topologies:

-- VL2~\cite{dc} is a 3-layer tree-like topology. Each top-of-rack (ToR) switch is connected with physical servers in the lower layer and aggregation switches in the upper layer. Aggregation switches are further connected to intermediate switches in the top layer. Bandwidths at different layers are set to 1Gbps, 10Gbps and 10Gbps, respectively. In a data center with $k$ aggregation switches, the number of intermediate switches is $\frac{k}{2}$; the number of ToR switches is $\frac{k^2}{4}$, each with 20 servers attached. 
We set $k=20$ to build a DL cluster with 2000 servers; there are 20 schedulers each managing 1 aggregation switch, 5 ToR switches and 100 servers. %Server architecture and workload are set as default.

-- BCube~\cite{dc} builds clusters in a recursive fashion. In a $k$-layer BCube topology, $BCube_0$ contains $n$ physical servers and one switch, and $BCube_k$ consists of $n$ $BCube_{k-1}$. We use $k=3$ and $n=6$ to construct a cluster with $n^{k+1}=1296$ servers. Every two $BCube_1$ are managed by a scheduler, and in total 18 schedulers are included. Bandwidths at different layers are set to 10Gbps, 20Gbps and 40Gbps, respectively.  %Server architecture and workload are set as default.

\begin{figure}[t]
\centering
\includegraphics[width=0.4\textwidth]{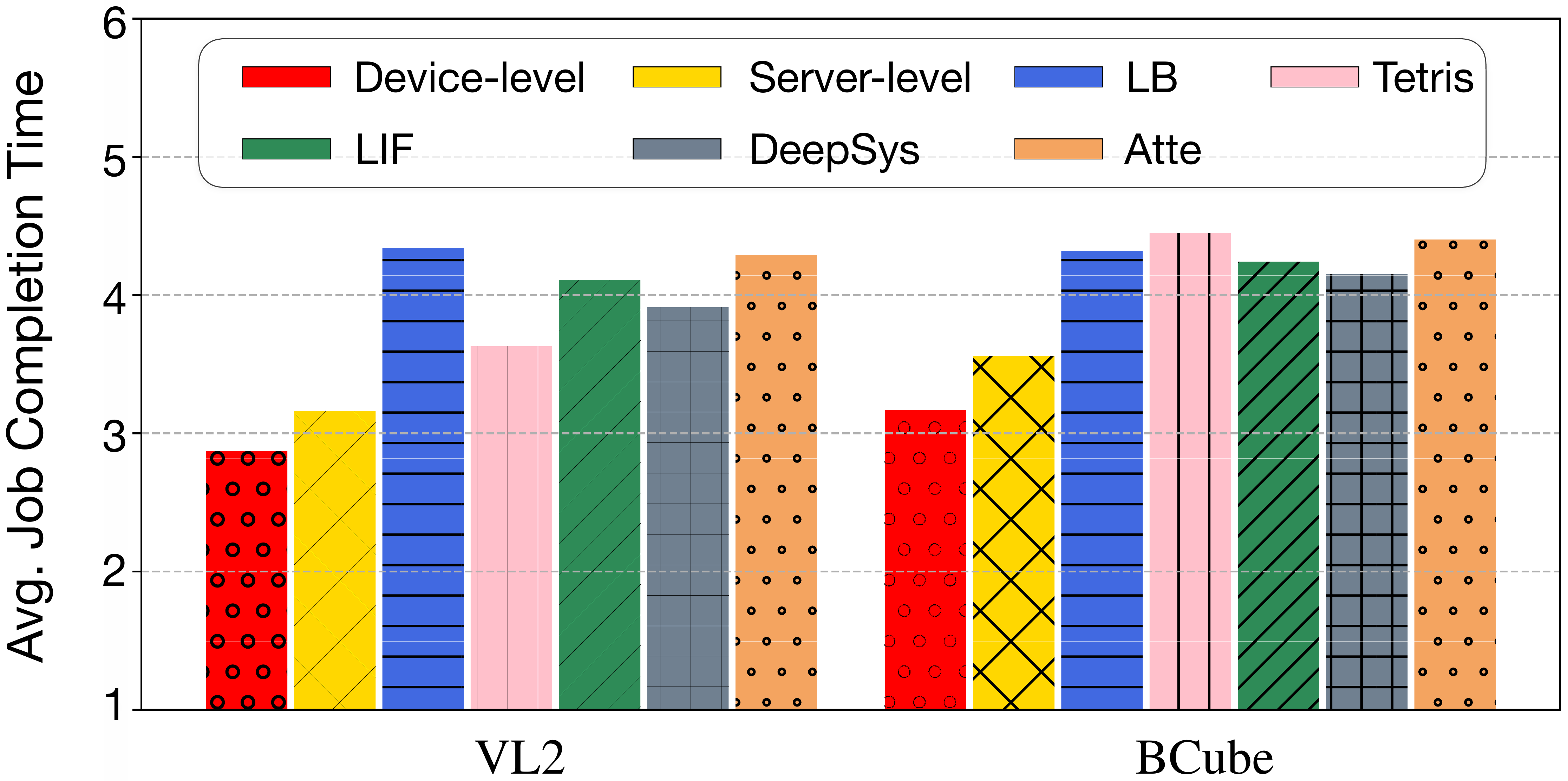}
\caption{Performance under different cluster topologies.}
\label{fig:exp-cluster}
\end{figure}

In Fig.~\ref{fig:exp-cluster},  %compares the performance of our framework with baselines when change fat-tree structure to VL2 and BCube respectively. 
we observe that our framework still outperforms baselines by at least 21\% under VL2 and BCube cluster topologies, further demonstrating the benefits of using graph embedding to generate topology-aware scheduling policy. 

\vspace{-3mm}
\subsection{Advantages of Multiple Agents}

We further compare the performance of using multiple schedulers with a single RL scheduler. When using one scheduler to schedule all jobs submitted to the whole cluster, we remove the generation of inter scheduler graph embedding from our proposed framework (i.e., the node feature after inner scheduler graph embedding will be fed into the DRL module as the input state) and the node set of the graph contains all GPU groups, CPUs and switches in the entire cluster.

\begin{figure}[t]
\centering
\includegraphics[width=0.485\textwidth]{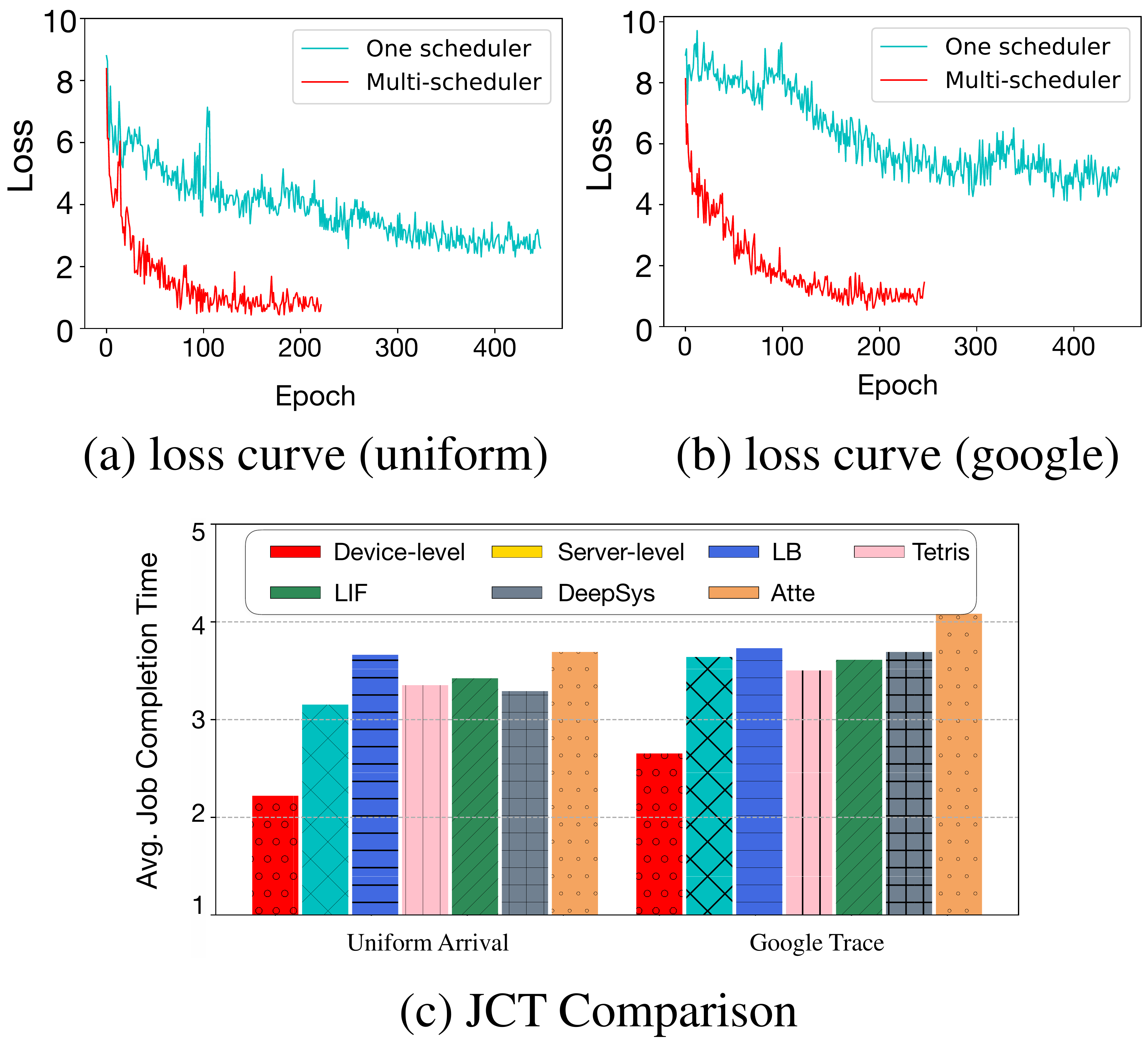}
\caption{Comparison between multiple schedulers (MARL) and one scheduler (RL) on the job set with Uniform arrival pattern and Google trace, in terms of convergence and job completion time.}
\label{fig:exp-multi-1}
\end{figure}

Fig.~\ref{fig:exp-multi-1} show the convergence curve and comparison of achieved average JCT, using uniform job arrival and Google trace, respectively. We can observe that using one scheduler to learn placement policy in a large-scale cluster is less efficient: (i) it takes more than 400 epochs to converge, while our framework with multiple schedulers only takes 200 epochs; (ii) it easily converges to a sub-optimal stage, which performs even worse than Tetris. One main issue for applying the RL is the inefficient exploration over a very large action space; what's worse, when the scale of the cluster increases, the single scheduler may not efficiently extract useful features about server status and workload placement because of the large input dimension. 
Using multiple schedulers with cooperative information exchange tackles these problems, thus achieving better convergence and placement policy.

In practice, using a single RL scheduler causes scheduling delay for newly arrived jobs, since it will perform multiple inferences to produce placement policies for jobs submitted to the whole cluster. Adopting multiple RL schedulers running on separate cluster partitions alleviates this burden.

% \begin{figure}[t]
% \centering
% \includegraphics[width=0.485\textwidth]{figures/exp-multi-1.pdf}
% \caption{Comparison between multiple schedulers (MARL) and one scheduler (RL) on the job set with Uniform arrival pattern, in terms of convergence and job completion time.}
% \label{fig:exp-multi-1}
% \end{figure}

% \begin{figure}[t]
% \centering
% \includegraphics[width=0.485\textwidth]{figures/exp-multi-2.pdf}
% \caption{Comparison between multiple schedulers (MARL) and one scheduler (RL) on the job set with Google trace, in terms of convergence and job completion time.}% and a customer of no one.}
% \label{fig:exp-multi-2}
% \end{figure}

\vspace{-3mm}
\subsection{Accuracy of Interference Model}

We compare the performance of our interference model with models as shown in Table~\ref{interference}. Linear (Quad) Model refers to using a linear (quadratic) function to model both CPU and PCIe interference, which is adopted in TRACON~\cite{tracon}, an interference-aware scheduler for virtualized applications. We also evaluate two variations of our interference model, neglecting PCIe interference (denoted as \textit{w/o PCIe}) and neglecting CPU contention (denoted as \textit{w/o CPU}). In total 480 samples with different job co-locations and corresponding slowdowns are collected from a server consisting of two Intel E5-2630 CPUs, four V100 GPUs and 128Gbps PCIe. We use the least-squares solver from $sklearn$~\cite{sklearn} to fit the models.

We observe that our model achieves less prediction error when estimating slowdown of co-located jobs, which helps the schedulers to efficiently learn interference-aware placement policy in offline training phase. Also, PCIe transmission and CPU utilization are two most important metrics to quantify resource contention for distributed DL jobs. Simply considering one of them fails to accurately measure the interference level. 

% ful linear, quad
\begin{table}[t]
\small
\centering
\caption{Comparison of prediction error using different fitting models.}\label{interference}
\begin{tabular}{|c|c|c|}
\hline
Fitting model   & Prediction error  \\ \hline
Linear Model (TRACON-Linear)    &   24.6\% \\
Quad Model (TRACON-Nonlinear)   &   22.9\% \\
Interference Model (ours)       &   \textbf{13.1\%} \\
Interference Model w/o PCIe     &   27.5\% \\
Interference Model w/o CPU      &   36.3\% \\ \hline

\end{tabular}
\end{table}

\section{Testbed Evaluation}

To further verify the effectiveness of our scheduling framework, we evaluate it using testbed experiments.

\vspace{-3mm}
\subsection{Implementation}

\noindent \textbf{Testbed.} We build a 48-GPU testbed on EKS, which is a managed service and certified Kubernetes conformant to run Kubernetes on AWS~\cite{eks}. The cluster has 12 \textit{g4dn.12xlarge} EC2 instances (i.e., servers), each configured as follows: 4 Nvidia T4 GPUs, 48 virtual CPU cores, 192 GB RAM, 50Gbps NIC, and PCIe supporting a throughput around 15GB/s.

\noindent \textbf{Workloads.} Jobs are submitted to the cluster with a down-scaled arrival rate extracted from the Google Trace. Each job randomly selects one model from the 8 types listed in Table~\ref{ml-table} as the training model. The number of PS required by one job is randomly set within $[1, 2]$; the number of workers required is randomly set within $[2, 4]$. Other job information, i.e., total number of epochs, resource demands, model size and mini-batch size, is set based on representative implementation~\cite{example}. PSs and workers are run in docker containers. Down-scaled training dataset is stored in servers in advance, and the file directory is then mounted to associated containers, avoiding downloading the dataset whenever a container is created.

\noindent \textbf{Schedulers on Kubernetes.} We implement our customized schedulers using Python on Kubernetes. There are 2 schedulers each managing 6 servers (24 GPUs) in the cluster. The scheduling interval is set to 10 minutes. At the beginning of a scheduling interval, each scheduler queries unscheduled jobs and current cluster status by sending HTTP requests to the Kubernetes API server and generates placement decisions through the DRL policy network. The schedulers will then deploy the PS or worker of each job as a separate Pod (i.e., a collection of containers sharing the storage and network namespace) to the selected server(s). Our schedulers are responsible for maintaining the life cycle of the Pods, including the Pod creating and Pod deleting. Schedulers update DRL models using online collected samples.

\begin{figure}[t]
\centering
\includegraphics[width=0.48\textwidth]{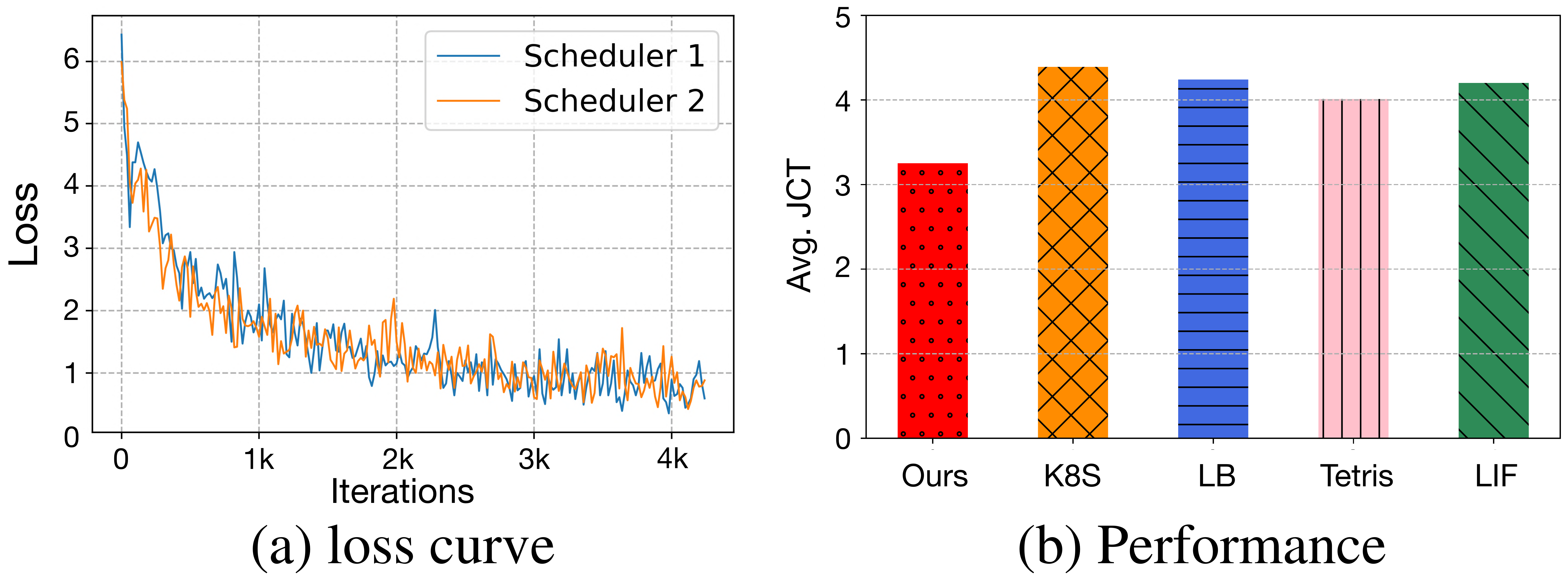}
\caption{(a) Loss curves of two schedulers. (b) Comparison among our framework and representative schemes.}
\label{fig:testbed-jct}
\end{figure}

\begin{figure}[t]
\centering
\includegraphics[width=0.47\textwidth]{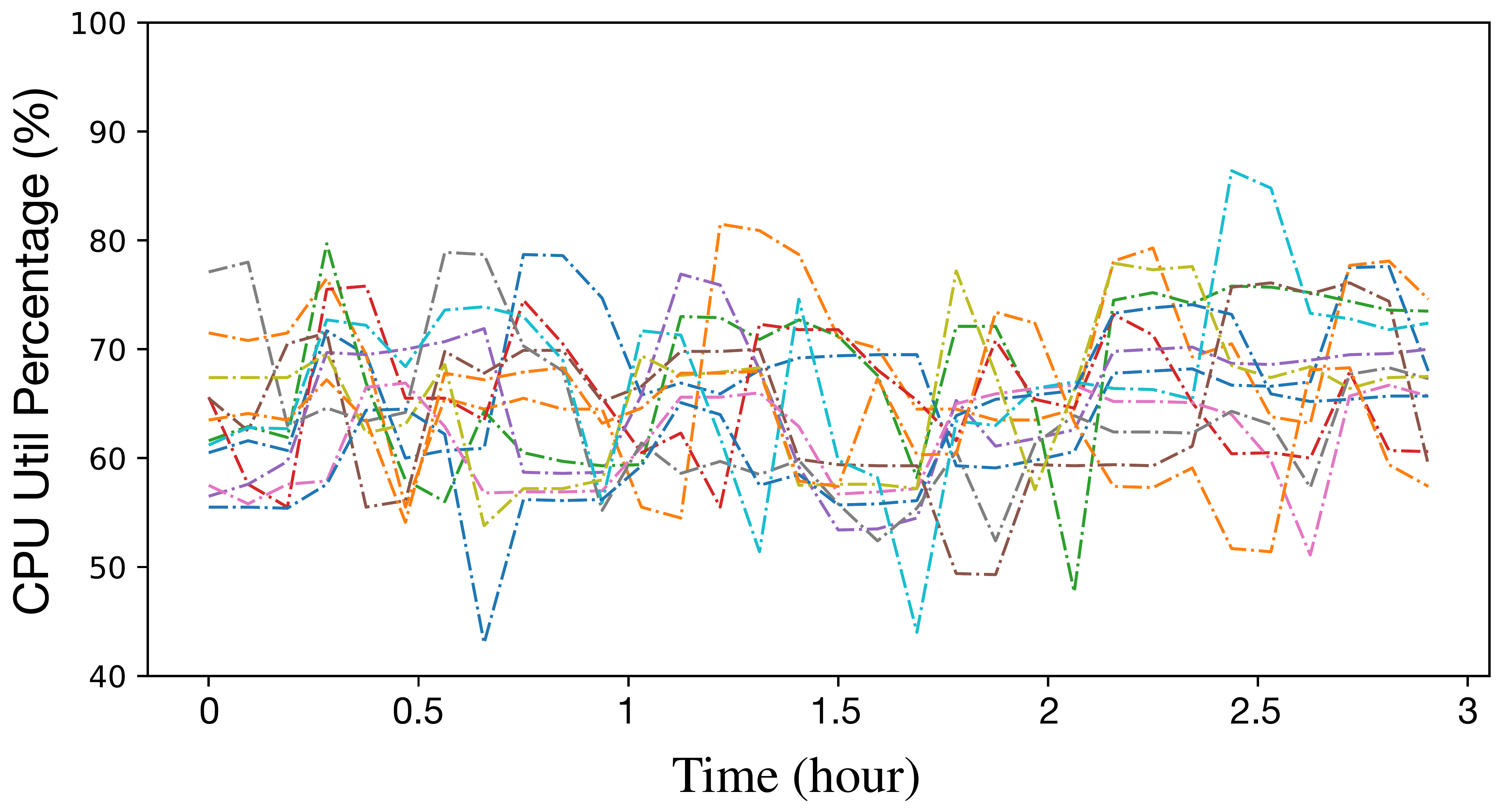}
\caption{CPU utilization profiled in a period of 3 hours: curves of different colors represent different servers.}
\label{fig:testbed-cpu}
\end{figure}

\vspace{-3mm}
\subsection{Performance}

We compare the performance of our schedulers in terms of JCT, as compared to Tetris, LB, LIF and Kubernetes' default placement (built-in load-balancer). Fig.~\ref{fig:testbed-jct}(a) shows the training loss of two schedulers. With more samples collected online, the loss decreases and converges gradually. In Fig.~\ref{fig:testbed-jct}(b), we can observe that our learnt scheduling policy achieves at least 18.9\% improvement as compared to baselines. It demonstrates the effectiveness of our scheduling framework in real environment. Due to resource limit, we did not have access to a testbed in the scale of thousands of servers. But our MARL framework is fully scalable, by introducing more schedulers as learning agents to manage more servers.

% handle complex env, restrict input dim

In Fig.~\ref{fig:testbed-cpu}, we also show the CPU utilization on all servers in a period of 3 hours, when the policy network becomes stable. Each server's CPU usage is profiled every 5 minutes with AWS Cloud Watch. We observe that the CPU utilization is controlled within a range (i.e., 55\% to 75\%) in every server. When a server has a high CPU usage, task with high demand of CPU tends not to be placed onto that server, in order to reduce the interference level. However, as compared to the purely load-balancing scheme, our schedulers take communication within one job into consideration at the same time.

% cpu utilization, load blanance, communication aware of the same job. 

\section{Related Work}
% 1 page
% MARL
% job scheduler in ML cluster 
% interference-aware task placement
\noindent \textbf{AI for Data Center Service.} 
% With the development of Artificial Intelligence (AI), 
Efforts have been devoted to applying AI in network management, services and systems. Especially in the context of data centers, AI has brought about substantial benefits to network service provisioning. For example, intelligent switches adopted in the AI Fabric project~\cite{aifabric} predict incoming traffic and enable higher network throughput; scheduling solutions~\cite{lbsurvey} using Recurrent Neural Network (RNN) achieve load balancing among servers in a data center; RL-based methods are adopted for power-efficiency data center management~\cite{power}, etc. We target job scheduling in ML clusters, which are more and more common in IT companies to support AI-driven services.

\vspace{1mm}
\noindent \textbf{ML Job Scheduler.} There are a number of recent studies on cluster scheduling. For improving resource utilization, Gandiva~\cite{gandiva} enhances efficiency of GPU by removing the exclusivity and fixed assignment of GPUs to DL jobs; AntMan~\cite{antman} introduces dynamic scaling mechanisms for both memory and computation, achieving GPU sharing between jobs and preventing interference. Themis~\cite{themis} addresses the unfairness of placement-sensitive characteristic in DL jobs by proposing a long term fairness object. Optimus~\cite{optimus} aims to reduce job completion time by allocating resources, with a speed fitting model. RL-based schedulers have also been designed. Decima~\cite{decima} learns workload-specific scheduling algorithms for Spark jobs. Placeto~\cite{placeto} learns a general device placement for operations through GNN embedding of a Spark job DAG. We focus on learning the interference-avoidance job placement to minimize job completion time in a large-scale ML cluster.

\vspace{1mm}
\noindent \textbf{Interference-aware Placement.} To mitigate performance degradation due to resource contention, Abhishek \textit{et al.}~\cite{hpc} propose an interference-aware VM placement policy, by characterizing applications based on their usage of shared resources. Bu \textit{et al.}~\cite{mapreduce} target building a task performance prediction model for MapReduce jobs, taking CPU usage and network I/O contention into consideration. Chiang \textit{et al.}~\cite{tracon} design an interference model to predict performance of VM applications, based on the observations of resource consumption. These white-box approaches rely on precise job profiling to determine coefficients in performance model. Harmony~\cite{harmony} and a DQN-based solution proposed by \textit{et al.}~\cite{springer-rl} equip a single scheduler with RL model, that encodes job interference implicitly to learn placement policy. Instead, to the best of our knowledge, we are the first to design an MARL framework with hierarchical GNNs to produce interference-aware and topology-aware placements.

% \vspace{1mm}
% \noindent \textbf{Topology-aware Scheduling.} 
% Faraji \textit{et al.}~\cite{topology16} propose a topology-aware GPU selection scheme to assign GPU devices to MPI processes. Amaral \textit{et al.}~\cite{topology17} schedule deep learning jobs on multi-GPU systems, taking different server architectures into consideration. Similarly, Lu \textit{et al.}~\cite{topology19} improve communication efficiency within one server, based on the number of concurrent workers and the topology of GPU placement. We aim to utilize the power of GNN to capture both intra-server topology and data center topology.

\vspace{1mm}
\noindent \textbf{Multi-agent Reinforcement Learning.}
Recently MARL has achieved promising results in various application domains, e.g., traffic engineering, video games, etc~\cite{marl}. We treat the DL cluster schedulers as a multi-agent system. The cooperation/communication among agents is important in MARL. We allow limited communication among neighboring schedulers, achieving sufficiently good respective field of each agent for placement decision making.
% MARL addresses the sequential decision-making problem of multiple autonomous agents that operate in a common environment, each of which aims to optimize its own long-term return by interacting with the environment and other agents~\cite{marl}. 

% ~\cite{CommNet}

\section{Conclusion}
% + ref 1 page
This paper proposes an MARL framework to schedule DL workloads in large-scale machine learning clusters. Multiple schedulers are adopted to manage partitions of the cluster, and they cooperatively make fine-grained placement decisions for distributed DL training jobs. Hierarchical GNNs with edge information are designed to encode the cluster topology and server architecture for topology-aware placement. 
Testbed and trace-driven evaluation shows that our schedulers efficiently learn good policies that mitigate job computation interference, and reduce gradient communication cost at the same time. Comparing with other representative scheduling schemes, at least 20\% improvement is achieved (in terms of JCT) under various settings. Further, our MARL framework achieves better policy convergence and higher learning speed than traditional RL deployed on a single scheduler, which implies its scalability in clusters with thousands of servers.

\ifCLASSOPTIONcaptionsoff
  \newpage
\fi

\bibliographystyle{IEEEtran}
\bibliography{IEEEabrv,reference}

\end{document}